\documentclass[12pt,preprint]{aastex}
\def\gs{\mathrel{\raise0.35ex\hbox{$\scriptstyle >$}\kern-0.6em \lower0.40ex\hbo
x{{$\scriptstyle \sim$}}}}
\def\ls{\mathrel{\raise0.35ex\hbox{$\scriptstyle <$}\kern-0.6em \lower0.40ex\hbo
x{{$\scriptstyle \sim$}}}}

\shorttitle{The Galaxy Luminosity Function in the Coma Cluster}
\shortauthors{Mobasher et al.}

\begin{document}
 
\title{A Photometric and Spectroscopic Study of Dwarf and Giant Galaxies 
in the Coma Cluster -  
IV. The Luminosity Function
\footnote{Based on observations made with the William Herschel Telescope 
operated on the island of La Palma by the Isaac Newton Group in the 
Spanish Observatorio del Roque de los Muchachos of the Instituto de 
Astrofisica de Canarias, and also the Anglo-Australian Telescope of the
Anglo-Australian Observatory.}}
\author{
Bahram Mobasher,$^{\!}$\altaffilmark{2,3}
Matthew Colless,$^{\!}$\altaffilmark{4}
Dave Carter,$^{\!}$\altaffilmark{5}
Bianca M.\ Poggianti,$^{\!}$\altaffilmark{6}
Terry J.\ Bridges,$^{\!}$\altaffilmark{7}
Kelly Kranz,$^{\!}$\altaffilmark{4}
Y. Komiyama,$^{\!}$\altaffilmark{8}
N. Kashikawa,$^{\!}$\altaffilmark{9}
M. Yagi,$^{\!}$\altaffilmark{9}
S. Okamura$^{\!}$\altaffilmark{10}
}
\smallskip

\affil{\scriptsize 2) Space Telescope Science Institute, 3700 San Martin 
Drive, Baltimore, MD 21218, USA}
\affil{\scriptsize 3) Affiliated to Space Telescope Division of the 
European Space Agency}
\affil{\scriptsize 4) Research School of Astronomy and Astrophysics, 
The Australian National University, Weston Creek, ACT 2611, Australia}
\affil{\scriptsize 5)  Astrophysics Research Institute, Liverpool John Moores
University, Twelve Quays House, Egerton Wharf, Birkenhead, Wirral, 
CH41 1LD, UK}
\affil{\scriptsize 6) Osservatorio Astronomico di Padova, 
vicolo dell'Osservatorio 5, 35122 Padova, Italy}
\affil{\scriptsize 7) Anglo-Australian Observatory, PO Box 296, Epping, 
NSW 2121, Australia}
\affil{\scriptsize 8) Subaru Telescope, 650 North Aohoku Place, Hilo, 
HI 96720}
\affil{\scriptsize 9) National Astronomical Observatory, Mitaka, Tokyo, 
181-8588, Japan}
\affil{\scriptsize 10) Department of Astronomy, University of Tokyo, 
Bunkyo-ku, Tokyo 113-0033, Japan}

\begin{abstract}
A large spectroscopic survey is constructed of galaxies in the Coma cluster. 
The survey covers a wide area (1 deg$^2$) to deep magnitudes ($R\sim 19.5$), 
covering both the core (high density) and outskirts (intermediate to low 
density) of the cluster. The spectroscopic sample consists of a total of 1191
galaxies, of which, 760 galaxies are confirmed members of the Coma cluster. 
A statistical technique is developed to correct the spectroscopic sample for 
incompletness. The corrected sample is then used to construct
R-band luminosity function (LF) spanning a range of 7 magnitudes 
($-23 < M_R - 5~log~h_{65} < -16$) both at the core and outskirts of 
the cluster. The R-band LF for the entire Coma cluster, fitted to Schechter
form, gives; $M^\ast_R = -21.79^{+0.08}_{-0.09} + 5~log~h_{65}$ and  
$\alpha = -1.18^{+0.04}_{-0.02}$. 

Dependence of the LF on local environment in Coma is explored. The LFs are
found to be the same, within the errors, between the inner and outer 
regions and close to those from recent measurements for field galaxies. 
This is remarkable given the variation in the spectral types of galaxies
between field and cluster environments. The steep faint-end slope 
for the LFs, 
observed in previous studies using photometric surveys, is not found here. 
However, the LF in this study is only measured to $M_R=-16$, compared to
much deeper limits ($M_R\sim -12$) achieved in photometric surveys. 

The total B-band LF for the Coma cluster, fitted to a Schecter form is;
$M^\ast_B = -19.95^{+0.15}_{-0.10} + 5~log~h_{65}$ and 
$\alpha = -0.96^{+0.01}_{-0.02}$. This also shows a dip at 
$M_B = -18$ mag., in agreement with previous studies. The implications
of this feature are discussed. 

The LF is studied in $B-R$ color intervals and shows a steep faint-end
slope for red ($B-R > 1.35$) galaxies, both at the core and outskirts of
the cluster. This population of low luminosity red galaxies
has a higher surface density than the  
blue ($B-R < 1.35$) star-forming population and dominates the faint-end of
the Coma cluster LF. 

It is found that relative number of high surface brightness
galaxies is larger at the cluster core, implying destruction
of low surface brightness galaxies in dense core environment.

\end{abstract}


\section{Introduction}

Detailed knowledge of the luminosity function (hereafter LF), defined as 
the number density of galaxies with a given luminosity, is essential for
any observational study of the formation and evolution of galaxies, and
for constraining galaxy formation scenarios and models for 
large scale structure. Over the last three decades, extensive studies of 
LFs have been performed, both in clusters and general fields. Although
the move from wide-area photographic plate surveys to high-performance,
sensitive CCDs has greatly advanced the subject, there are still a number
of questions unanswered, including:

\begin{itemize}

\item Is a single parametric form for the LF a reasonable 
approximation over the entire luminosity range?

\item What is the dependence of the LF on environment and does this also depend on
the selection characteristics of the sample (i.e.\ depth, completeness)?

\item How do the LFs for different Hubble types or color intervals
compare?

\item Do the high and low surface brightness galaxies follow different LFs? 

\item How do the LFs of giant and dwarf galaxies 
compare?

\end{itemize}

To address these questions, one needs a wide-area survey, complete to deep
magnitudes. Moreover, in comparing different LFs, one must minimise
observational biases (i.e.\ different depths) and selection effects 
(i.e.\ incompleteness). These are some of the reasons for 
slow progress in this field. 

Study of the field LF requires a redshift survey, complete to some
magnitude limit. Due to the small numbers of intrinsically faint galaxies 
in even the largest magnitude-limited redshift surveys, field LFs do not
normally extend to faint magnitudes.
While this can be avoided for cluster galaxies, as all the objects are 
roughly at the same distance, there are still problems at faint magnitudes 
due to contamination by background objects. Moreover, 
compared to isolated systems, 
cluster galaxies are subject to dynamical
evolution, affecting the shape and characteristics of their LF. 
The former problem can be resolved by exploiting a deep spectroscopic
sample in clusters (and so spectroscopically-confirmed cluster members), 
while the latter is best addressed by studying large areas around clusters, 
covering both low and high-density regions. 

In this paper, we aim to explore the above questions by constructing
the LF in the Coma cluster, avoiding potential problems which have affected 
previous studies. This has become
possible with the advent of large-format CCDs, allowing wide-area surveys
to deep magnitudes. Combined with fiber-fed spectrographs, one could then
construct statistical samples of spectroscopic data on galaxies to faint 
magnitudes. 

At a redshift of $z=0.023$, Coma is the richest nearby cluster, 
providing a laboratory for studying evolution of different types
of galaxies. Other nearby rich clusters, such as Virgo and
Ursa Major, are relatively unevolved, not allowing a direct 
comparison of the properties of galaxies as a function of their 
local environment. A detailed study of Coma also provides a
control sample to compare with more distant clusters. 
This improves previous works in many aspects, including:

\begin{itemize}

\item wide-area CCD coverage of the cluster, extending to $>$1\,deg. from
its center (this also covers the NGC4839 group);

\item full spectroscopic information for galaxies in the sample,
allowing a measure of the LF independent of uncertain background
corrections;

\item inclusion of the faintest galaxies for which spectroscopic data
can be obtained (R$\sim$19.5), allowing strong constraints
on the shape of the LF at faint magnitudes;

\item availability of accurate CCD photometry in both $B$ and $R$ bands, 
providing color information for galaxies in the sample;

\item well-defined selection function for the spectroscopic sample. 

\end{itemize}

This is the fourth paper in a series studying the nature of giant and dwarf galaxies
in the Coma cluster, using photometric (Komiyama et al.\ 2001; paper I) and 
spectroscopic (Mobasher et al.\ 2001; paper II) observations. Analyses of the
diagnostic line indices (Poggianti et al.\ 2001; paper III), the radial 
dependence of galaxy properties (Carter et al.\ 2002; paper V) and the
dynamics of giant and dwarf populations (Edwards et al.\ 2002) are already
performed. Future papers will present a spectroscopic comparison with
intermediate redshift clusters (Pogiantti et al.\ 2002; in prep.) and a study of
the scatter and environmental dependence of the color-magnitude relation 
(Mobasher et al.\ 2002; in prep.).  

In the next section, observations and spectroscopic sample selection are 
discussed. The selection functions are derived in section 3. Section 4
presents the LFs and their dependence on different physical parameters. 
The results are then compared with other studies in section 5,  
followed by a discussion of the results in section 6. 
The conclusions are listed in section 7.  
We assume a distance modulus of $DM =35.13$ mag. for the Coma cluster, 
corresponding to $H_0=65$ km/sec/Mpc. 

\section {Observations and Sample Selection}

To address the aims of this study, photometric and spectroscopic
observations were designed to survey a wide area of the cluster (consisting of
both core and outskirt fields) to deep flux levels. This is then
complemented by a sample of brighter galaxies with spectroscopic data, 
compiled from different sources. The result is a sufficiently large
spectroscopic sample to allow a detailed study of the LF of galaxies. 
The procedure for selecting the 
spectroscopic sample and its properties are summarised as follows.

A wide-area photometric survey was carried out by the MCCD (a mosaic camera
consisting of $5\times8$ CCDs and a field of view of 0.5\,deg$^2$; Sekiguchi
et al.\ 1998)
on the WHT. A total of five fields in the Coma
cluster were surveyed, each $32.5\times50.8$\,arcmin$^2$. 
The photometric surveys were performed
in both $B$ and $R$ bands, and are complete to $R=21$. Details of
the photometric observations, data reductions, source
extraction and the construction of the photometric catalogs 
are given in Komiyama et al.\ (2001; paper I). 
Follow-up medium resolution (6--9\AA) spectroscopic observations, using WYFFOS on the WHT,
were then performed on two fields of this survey: Coma1 (at the center)
and Coma3 (south-west of Coma1, including the NGC4839 group). 
The spectroscopic sample has a 
magnitude limit of $R$=19.75, with the spectra having large enough S/N ratios
to allow accurate measurement of 
diagnostic line indices (Poggianti et al.\ 2001; 
Paper III). 
The coordinates of the center of fields and the 
total number of objects in the spectroscopic sample in each field is 
listed in Table~1. The spectroscopic sample selection, spectroscopic 
observations 
and data reduction are explained in detail in Mobasher et al.\ 
(2001; paper II), where the spectroscopic catalog is also presented. 
The criteria for selecting the spectroscopic sample in the central (Coma1)
and outskirt (Coma3) fields are the same, with no radially-dependent 
selection biases present. This sample will be refered to as
the Deep Spectroscopic Survey (DSS).

\begin{table*}
\caption{Coordinates of the center of each field and the number of sources 
with spectroscopic data in the SSS
and DSS samples\vspace{12pt}}
\begin{tabular}{cccrrrr}
Field & R.A.       & Dec.         & \multicolumn{2}{c}{DSS} & \multicolumn{2}{c}{SSS} \\
      & \multicolumn{2}{c}{J2000} & Members      & Total    &  Members      & Total   \\
Coma1 & 12~59~23.7 & 28~01~12.5   & 189          & 302      &  282          & 369     \\
Coma2 & 12~57~07.5 & 28~01~12.5   & ---          & ---      &   97          & 184     \\
Coma3 & 12~57~07.5 & 27~11~13.0   &  90          & 188      &  102          & 148     \\
\end{tabular}
\end{table*}

The DSS is complemented by another sample with 
a brighter spectroscopic magnitude limit ($R\sim18$), based 
on the compilation from Colless \& Dunn (1996) and another  
spectroscopic survey of Coma galaxies, carried
out using the 2-degree Field (2dF)
spectroscopic facility at the AAT (Edwards et al.\ 2002).  
This sample is refered to as the Shallow Spectroscopic Survey (SSS). The total
number of galaxies in each field for the SSS sample is also listed in Table~1. 
Line indices are not available for the SSS sample.

The spatial distribution of galaxies in the DSS and SSS are
shown in Figure~1, with the location and size of the MCCD survey areas 
overlaid: the Coma1 field at the center of the cluster, 
the Coma2 field to the west of Coma1, and the Coma3 field to the south-west of
Coma1. 
Comparison between the redshifts of galaxies common to both the DSS and SSS
samples shows a mean difference of $28\pm$9\,km\,s$^{-1}$ 
between the estimated velocities in the two surveys.

\begin{figure}
\plotone{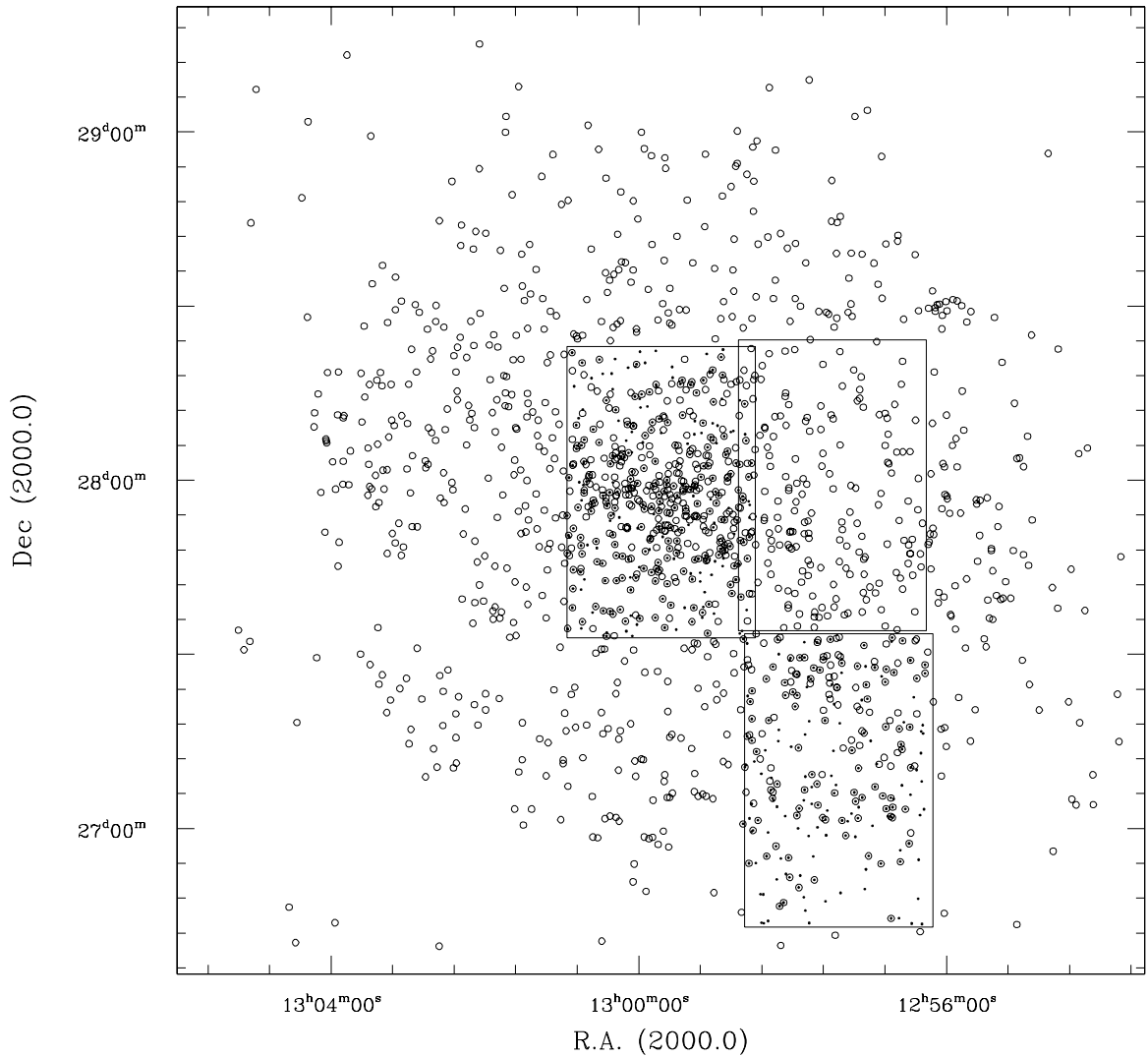}
\caption{The distribution of galaxies with spectroscopic data from DSS
(solid points); SSS (open circles) and galaxies in common to both surveys
(dotted circles).  
The three fields with available spectroscopic data are shown 
as rectangles: Coma1 (central), 
Coma2 (west of Coma1) and Coma3 (south-west of Coma1).
The total number of galaxies and the
coordinates of the center of each field are listed in Table~1.
\label{fig1}}
\end{figure}

The final spectroscopic catalog is compiled by combining galaxies with
available redshifts from the DSS and SSS samples, the $B$ and $R$-band CCD
data (i.e.\ magnitudes and surface brightness values) from the MCCD
photometric survey, and the line indices for galaxies in the Coma1 and Coma3 fields.  
For galaxies common to both spectroscopic
surveys, the mean redshifts were calculated. The $B$ and $R$-band magnitudes 
in this study are measured over a circular aperture of radius 3 times the Kron
radius (see Paper II). 
This has the advantage of scaling the photometric 
aperture in proportion to the size of galaxies
and is therefore less susceptible to under-estimating the magnitudes for larger
galaxies, as is often the case if a constant aperture is used for all
the objects at the same distance. The Kron magnitudes used here are
close to total (see paper I for details).   
The $B-R$ colors are also measured over circular 
apertures of 3 times the Kron radius and hence, 
correspond closely to total colors. 

Galaxies in the range $4000 < cz < 10000$\,km\,s$^{-1}$ are considered to be members of 
the Coma cluster (Colless \& Dunn 1996). 
The total number of galaxies identified as cluster members in each of 
the three fields here, are also listed in Table 1. 
In case of Coma2 field, only spectroscopic data from the SSS
are available. For this reason, the depth of the spectroscopic survey in this
field is shallower than that in the other two fields, with 
a different completeness function, as discussed in the next section.

The photometric properties of galaxies in the spectroscopic sample are
explored using their $B-R$ colors and $R$-band effective surface brightness 
($\mu_{\rm eff}$) distributions in Figures~2 and~3 respectively. 
The effective surface brightness is defined as the
mean surface brightness within the effective radius (radius containing
half the total light of the galaxy, with the latter assumed to be the 
luminosity
inside an aperure 3 times the Kron radius of that galaxy- paper I) of galaxy. 
The effect of seeing on the observed surface brightness was considered and 
found to be negligible. 

There are 
four galaxies with $B-R>2$ which are confirmed members of the Coma cluster
(2 galaxies in Coma1 and one galaxy in each of Coma2 and Coma3 fields).
Such red galaxies were previously considered to be background objects 
(Conselice et al 2002), with none so far identified as a Coma cluster member. 
Considering the surface brightness distributions, we find that cluster members
have a relatively brighter effective surface brightness 
compared to non-members (when corrected for surface brightness dimming 
and the K-correction due to
redshift), with the tail of the surface brightness
distribution for cluster members extending to brighter values. The median
effective surface brightness changes from $\mu_{eff} = 20.50$ mag/arcsec$^2$
in the cluster to 21.50 mag/arcsec$^2$ for the field galaxies. 
While this was known
from previous studies, there is also evidence for a monotonic decrease in 
surface brightness from core (Coma1) to outskirts (Coma3). 

\begin{figure}
\plotone{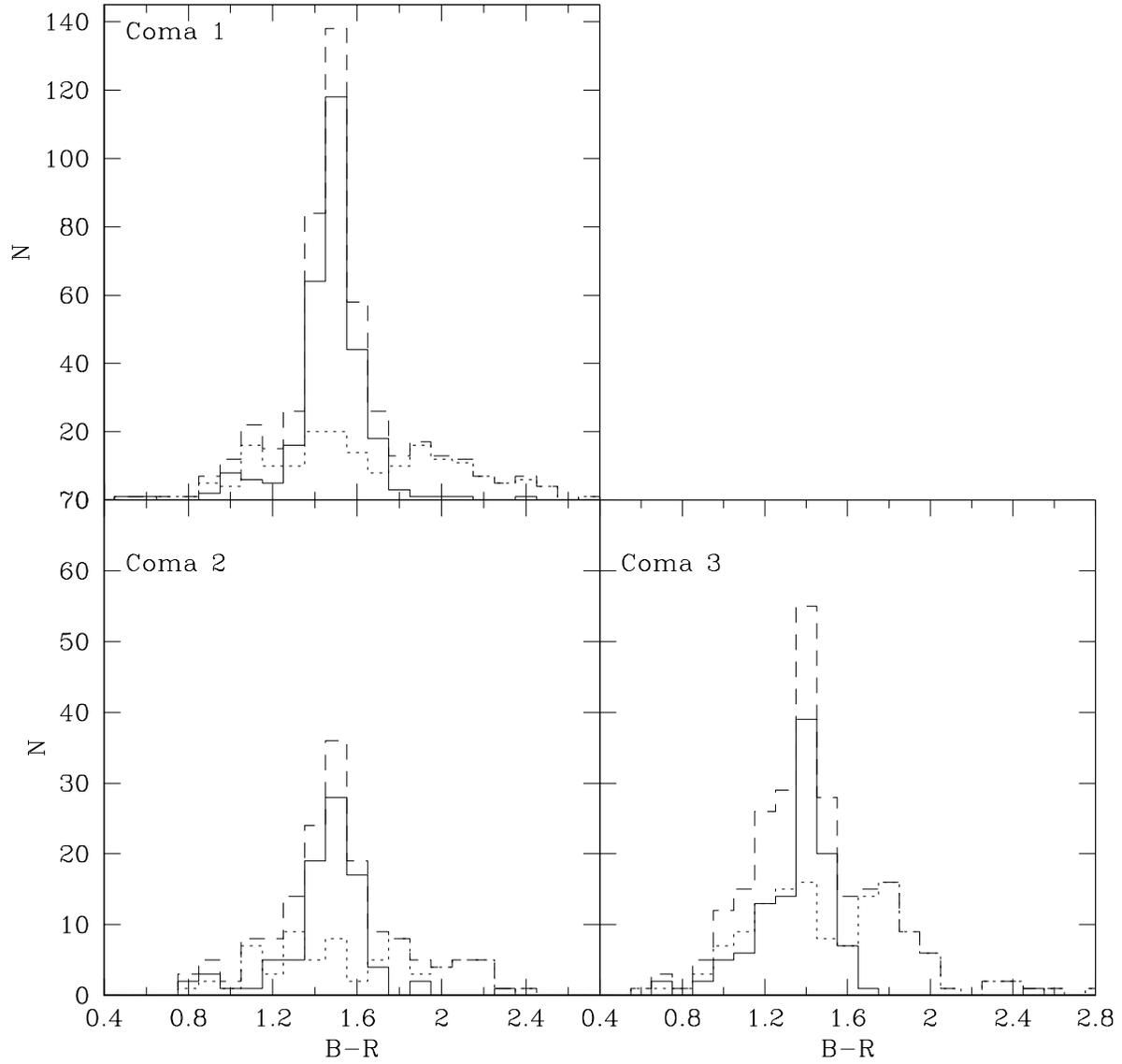}
\caption{$B-R$ color distributions in the three fields for the entire
DSS+SSS spectroscopic sample (dashed line), for
spectroscopically-confirmed cluster members 
(solid line), and for non-members (dotted line).
\label{fig2}}
\end{figure}

\begin{figure}
\plotone{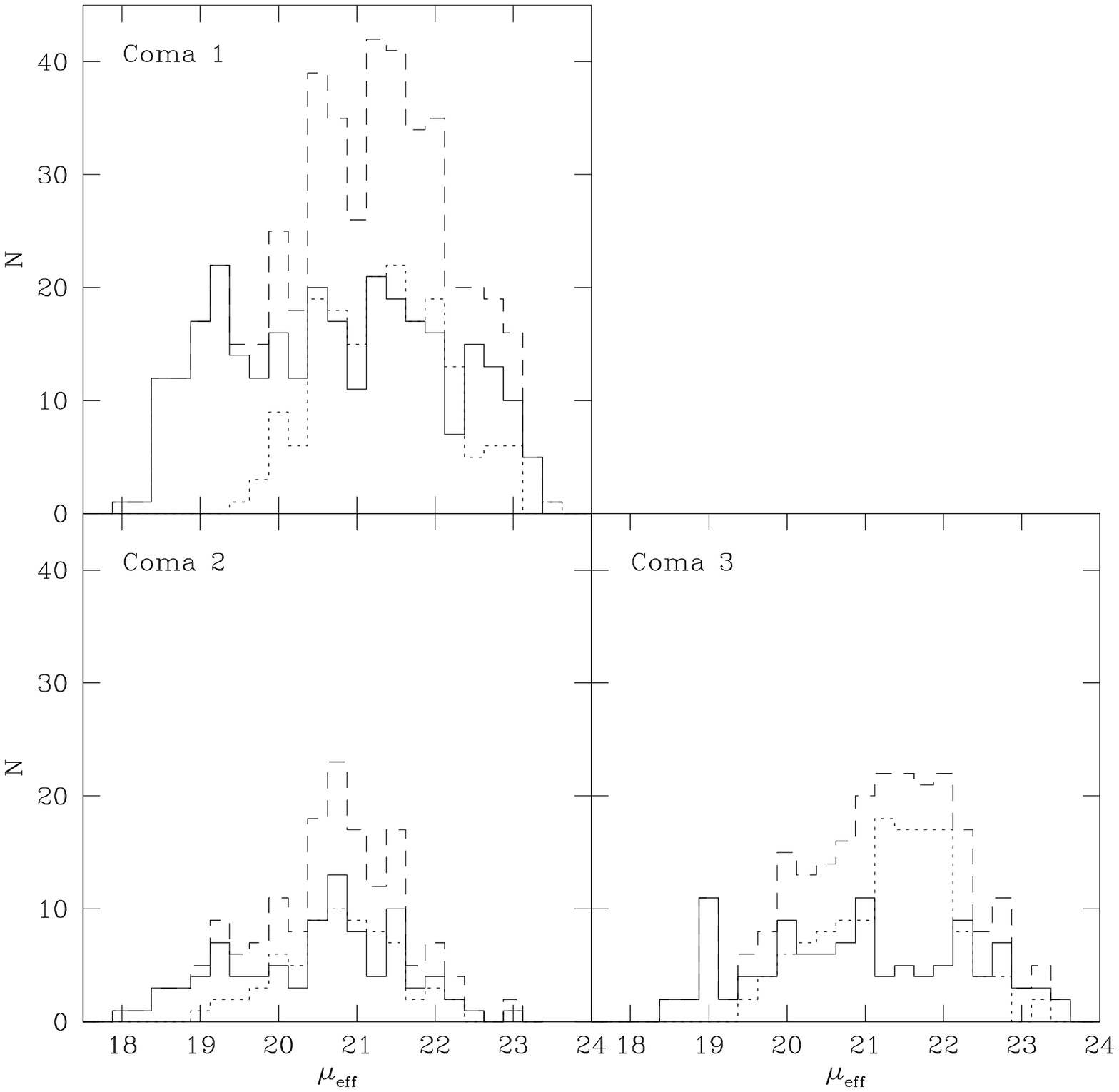}
\caption{$R$-band surface brightness distributions in the three fields
for the entire DSS+SSS
spectroscopic sample (dashed line), for
spectroscopically-confirmed cluster members 
(solid line), and for non-members (dotted line).
\label{fig3}}
\end{figure}

\section {Cluster membership}

The spectroscopic surveys are incomplete, so we do not know for every
individual galaxy whether or not it is a member of the cluster.
Therefore, in order to compute the cluster LF, we
assume that the spectroscopic sample is representative, in the sense
that the fraction of galaxies that are cluster members (which depends on
both magnitude and position) is the same in the (incomplete)
spectroscopic sample as in the (complete) photometric sample. If this
assumption applies to both the spectroscopic surveys we are using, then
they can be straightforwardly combined to give the cluster membership
fraction over the full magnitude range of the two surveys.

If $N_{c}(m,r|{\rm spec})$ and $N_{t}(m,r|{\rm spec})$ are,
respectively, the number of cluster members (based on their redshifts)
and the total number of galaxies within the combined spectroscopic sample
in given ranges of apparent magnitude and cluster radius, then the
probability of a randomly-selected galaxy in the combined spectroscopic
sample being a member of the Coma cluster, $P(m,r|{\rm spec})$, is

$$P(m,r|{\rm spec}) = { N_{c}(m,r|{\rm spec}) \over N_{t}(m,r|{\rm spec}) } .$$

\noindent We simplify this by taking as our radial bins the ranges
covered by the three fields of the survey, so that we have

$$ P_i(m|{\rm spec}) = { N_{c,i}(m|{\rm spec}) \over N_{t,i}(m|{\rm spec}) } ,$$

\noindent where $i=1,2,3$ corresponds to the Coma1, Coma2 and Coma3
fields.

The cluster membership fractions as functions of apparent magnitude are
presented in Figure~4 and listed in Table 2 for each of the three fields. 
The uncertainties in these fractions, assuming Poisson statistics, 
are estimated as

$${dP_i(m|{\rm spec})\over P_i(m|{\rm spec})} = ({1\over N_{c,i} 
(m|{\rm spec})} - {1\over N_{t,i} (m|{\rm spec})})^{1/2}$$

\noindent where $N_{c,i} (m|{\rm spec})$ is a binomial variable with its
corresponding variance, as discussed in the next section. 
The errors corresponding to each magnitude interval in each
field are also listed in Table 2. The membership
fraction is determined down to $R\sim 19.5$ for the Coma1 and Coma3
fields, but only down to $R$=17.75 for Coma2 (since only the SSS is
available in this field). As expected, the Coma1 field, being centrally
located, has a higher membership fraction at all magnitudes than the
Coma2 and Coma3 fields. Coma3 has a comparable membership fraction to
Coma2 at all magnitudes, even though it is further from the centre of
the cluster; this reflects the presence of the NGC4839 group in this
field.

\begin{table*}
\caption{Membership fractions in R-band magnitude intervals
for the three fields. Errors are given in brackets. Photometric errors
(0.03 mag) are significantly smaller than the bin sizes \vspace {12pt}}
\begin{tabular}{cccccc}
R     & Coma1        & & Coma2        & & Coma3 \\
      &              & &              & &       \\
12.75 & 1.00\ (0.58) & & $-$          & &   1.00\ (0.71) \\  
13.25 & 1.00\ (0.27) & & 1.00\ (0.58) & &   1.00\ (1.00) \\
13.75 & 1.00\ (0.27) & & 1.00\ (0.45) & &   1.00\ (0.44) \\
14.25 & 1.00\ (0.20) & & 0.92\ (0.28) & & 0.89\ (0.31) \\
14.75 & 0.96\ (0.18) &  & 0.75\ (0.43) & & 0.93\ (0.25) \\
15.25 & 0.95\ (0.22) &  & 0.83\ (0.26) & & 0.80\ (0.40) \\
15.75 & 0.86\ (0.17) &  & 0.79\ (0.20) & & 0.73\ (0.26) \\ 
16.25 & 0.94\ (0.16) &  & 0.60\ (0.17) & & 0.77\ (0.24) \\
16.75 & 0.66\ (0.11) &  & 0.57\ (0.14) & & 0.36\ (0.13) \\
17.25 & 0.59\ (0.01) &  & 0.35\ (0.01) & & 0.59\ (0.16) \\
17.75 & 0.42\ (0.08) &  & 0.29\ (0.01) & & 0.40\ (0.01) \\
18.25 & 0.42\ (0.09) &  & $-$             & & 0.23\ (0.09) \\
18.75 & 0.33\ (0.08) &  & $-$             & & 0.32\ (0.09) \\
19.25 & 0.23\ (0.09) &  & $-$             & & 0.07\ (0.05) \\
19.75 & $-$          &  & $-$             & & 0.25\ (0.25) \\
\end{tabular}
\end{table*}

\begin{figure}
\plotone{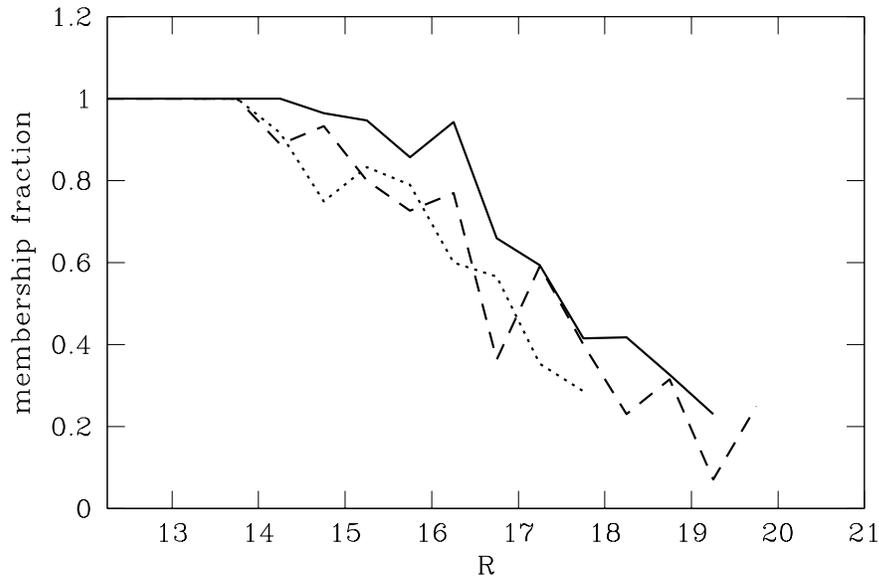}
\caption{Cluster membership fractions as a function of apparent
magnitude for Coma1 (solid line), Coma2 (dotted line) and Coma3 (dashed
line).
\label{fig4}}
\end{figure}

\section {The cluster luminosity function}

Let $N_{p,i}(m)$ be the number of galaxies as a function
of magnitude $m$ and field $i$, in the full photometric sample. 
The cluster LF can then be estimated as
\begin{equation}
\phi_i(m) = \frac{N_{p,i}(m)}{A_i}\, f_i(m)~~,
\end{equation}
where $A_i$ is the area of each field,  
$N_{p,i}(m)/A_i$ the surface density in the photometric
catalogue, and $f_i(m)= P_i(m|{\rm spec})$ is the cluster membership fraction.

The errors in the LF can be estimated by noting that $\phi_i(m)$ is the
product of two random variables: $N_{p,i}(m)$, which is a Poisson random
variable, and $N_{c,i}(m|spec)$, which is a binomial random variable (since it
is the number of cluster members found in a fixed number of trials,
$N_{t,i}(m|spec)$, where the probability of a single trial yielding a cluster
member is the true cluster membership fraction, $f_i(m)$). The relative
error in the LF is given by the quadrature sum of the relative errors in
$N_{p,i}(m)$ and $N_{c,i}(m|spec)$, so that
\begin{equation}
\frac{{\rm var}[\phi_i(m)]}{\phi_i(m)^2} =
       \frac{{\rm var}[N_{p,i}(m)]}{N_{p,i}(m)^2} + 
       \frac{{\rm var}[N_{c,i}(m|spec)]}{N_{c,i}(m|spec)^2} ~~.
\end{equation}
The variance in the Poisson variable is its expectation value, 
${\rm var}[N_{p,i}(m)]=N_{p,i}(m)$, while the variance in the binomial
variable is ${\rm var}[N_{c,i}(m|spec)]=N_{t,i}(m|spec))f_i(m)[1-f_i(m)]$. 
The relative error in the LF is thus 
\begin{equation}
\frac{\delta\phi_i(m)}{\phi_i(m)} \equiv
\frac{{\rm var}[\phi_i(m)]^{1/2}}{\phi_i(m)} =
\left( \frac{1}{N_{p,i}(m)} + \frac{1}{N_{c,i}(m|spec)} -
\frac{1}{N_{t,i}(m|spec)}\right)^{1/2} ~~.
\end{equation}
This expression reduces to $N_{c,i}(m|spec)^{-1/2}$,  
if $N_{t,i}(m|spec)=N_{p,i}(m)$
(i.e.\ if the spectroscopic sample is complete), and to $N_{p,i}(m)^{-1/2}$, 
if $N_{c,i}(m|spec)=N_{t,i}(m|spec)$ (i.e.\ if the membership fraction 
is unity).

The LFs for the Coma1, Coma2 and Coma3 fields are listed in Table~3
and presented in Figure~5. Magnitudes are corrected for both Galactic 
absorption ($A_R = 0.02$) and redshift ($K_z = 0.02$) dimming. The
distance of the Coma cluster is then estimated as $m-M = DM + A_R + K_z$
where $DM = 35.13$ mag. is its distance modulus (corresponding to
$H_0 =65$ km/sec/Mpc). At this distance, 
$1^\circ = 1.77$ Mpc (assuming $q_0=0.5$), corresponding to an area of 
1.619 Mpc$^2$ for each of the three fields surveyed in this study. The LFs
span the range $-22 < M_R - 5\ log\ h_{65} < -16$, except for the Coma2 field, which is only
based on the shallower SSS sample and hence, only extends to $M_R = -18$.

\begin{table*}
\caption{The observed LFs and their associated errors for the three fields. 
LFs are in galaxies/mag/Mpc$^2$. Galaxies from both the extended SSS 
(lines 1-9) and DSS+SSS (lines 10-24) are used. A distance modulus of 
$DM =35.13$ ($H_0=65$ Km/sec/Mpc) is assumed for Coma. 
\vspace{12pt}}
\begin{tabular}{ccccccc}
 & \multicolumn{2}{c}{Coma1} & \multicolumn{2}{c}{Coma2} & \multicolumn{2}{c}{Coma3} \\
$M_R$ & $\phi_1(M)$ & $\Delta(\phi_1(M))$ & $\phi_2(M)$ & $\Delta(\phi_2(M))$ &
$\phi_3(M)$ & $\Delta(\phi_3(M))$ \\
        &      &      &      &      &      &     \\
\multicolumn{3}{c}{Extended SSS}       &      &      &      &     \\
        &      &      &      &      &      &     \\
 -23.00 & 2.16 & 1.53 & 0.72 & 0.51 & 0.72 & 0.51 \\
 -22.75 & 3.24 & 1.87 & 1.08 & 0.62 & 1.08 & 0.62  \\
 -22.50 & 3.24 & 1.87 & 1.08  & 0.62 & 1.08  & 0.62  \\
 -22.25 & 5.41 & 2.42 & 1.80  & 0.81 &  1.80 &  0.81 \\
 -22.00 & 14.06 & 3.90 & 4.69  & 1.30  &  4.69 &   1.30  \\
 -21.75 & 12.98 & 3.75 & 4.33  & 1.25  &  4.33 &   1.25  \\
 -21.50 & 21.63 & 4.84 & 7.21  & 1.61  &  7.21 &   1.61  \\
 -21.25 & 24.87 & 5.19 & 8.29  & 1.73  &  8.29 &   1.73  \\
 -21.00 & 21.63 & 4.84 & 7.21  & 1.61  &  7.21 &   1.61  \\
        &      &      &      &      &      &     \\
\multicolumn{3}{c}{DSS + SSS}       &      &      &      &     \\
        &      &      &      &      &      &     \\
 -22.50 & 3.71 & 2.14 &   $-$       &  $-$        &  2.47 &   1.75 \\  
 -22.00 & 13.59 & 4.10 & 3.71  & 2.14  &  2.47 &   1.75 \\
 -21.50 & 18.53 & 4.78 & 4.94  & 2.47  &  2.47 &   1.75 \\
 -21.00 & 24.71 & 5.52 & 7.93  & 3.07  &  10.98 &   3.71 \\
 -20.50 & 26.21 & 5.67 & 11.12  & 4.54  &  16.14 &   4.45 \\
 -20.00 & 39.79 & 7.16 & 10.29 & 3.52  &  9.88 &   3.83 \\
 -19.50 & 26.47 & 5.67 & 15.60  & 4.32  &  10.78 &   3.69 \\
 -19.00 & 39.60 & 6.99 & 14.82  & 4.28  &  13.30 &   4.09 \\
 -18.50 & 33.41 & 6.28 & 24.51  & 5.70  &  8.086 &   2.97 \\
 -18.00 & 50.56 & 8.17 & 18.75  & 5.21  &  21.17 &   5.44 \\
 -17.50 & 46.18 & 8.36 & 22.94  & 7.42  &  27.67 &   6.51 \\
 -17.00 & 62.51 & 11.45 &   $-$       &  $-$        &  28.51 &   10.60 \\
 -16.50 & 82.07 & 16.88 &   $-$       &     $-$     &  59.30 &   14.95 \\
 -16.00 & 80.39 & 29.18 &   $-$       &  $-$        &  18.35 &   12.57 \\
 -15.50   &  $-$       &   $-$  &   $-$ & $-$   &  104.08 &  90.31 \\
\end{tabular}
\end{table*}

It is clear from Figure~5 that the bright-end of the LFs are not well
constrained from these three 
fields (filled circles). This leads to instability in 
parametric fits to LFs, as the characteristic magnitudes and faint-end
slopes are known to be correlated (see the next section for details). To 
overcome this, we construct the bright-end of the LFs by combining the present
survey (DSS+SSS) with an extension of the SSS sample, covering
a larger area of the Coma cluster. 
This extended survey covers an area of $1^\circ$ radius (corresponding
to the field of view of 2dF), centered on the Coma core, and to the same
depth as the SSS sample, substantially increasing the number of 
bright galaxies in the spectroscopic Coma survey. Here, we assume that
the shape of the bright-end of the LFs is independent from their local
environment, with the implication of this assumption discussed in 
section 4.1. However, photometry for 
galaxies in the extended SSS sample 
is only available in $b$ and $r$ bands, measured from photographic 
plates (Godwin et al 1983). To convert these to 
the photometric system used in this study, we use galaxies with available
photometry in both systems and derive the empirical relation

$$R = r + k_1 (b-r) + k_2$$

\noindent where $k_1 = 0.101\pm 0.068$ and $k_2 = 0.143 \pm 0.037$ with 
$rms= 0.218$ mag. This relation is subsequently used to convert magnitudes
in the extended SSS sample to those of the MCCD photometric system of the
present study. 

The membership fraction is estimated for this sample, as discussed in the
last section, and used to construct the LF over a 1-degree radius region
about the cluster center. The LF, 
binned in 0.25 mag. intervals, is then normalised to those 
of Coma1, Coma2 and Coma3 fields, derived from the DSS and SSS samples, 
using galaxies with $M_R < -21$ mag., and shown by open circles in Figure 5.
This extends the bright-end of the LFs to $M_R =-23$. By 
adopting the extended SSS sample here, we impose strong constraints on the
bright-end of the LF for each field while, at the same time, retaining 
the main advantage of
the present sample in extending the Coma LF to the faintest possible 
magnitudes. The estimated bright-end of the LF, derived from the 
extended sample and normalised to the LFs in each of the three fields, are 
also listed in Table 3. 
In the next section, parametric fits are carried out to the
observed LFs over the entire magnitude range shown in Figure 5.

\begin{figure}
\figurenum{5}
\plotone{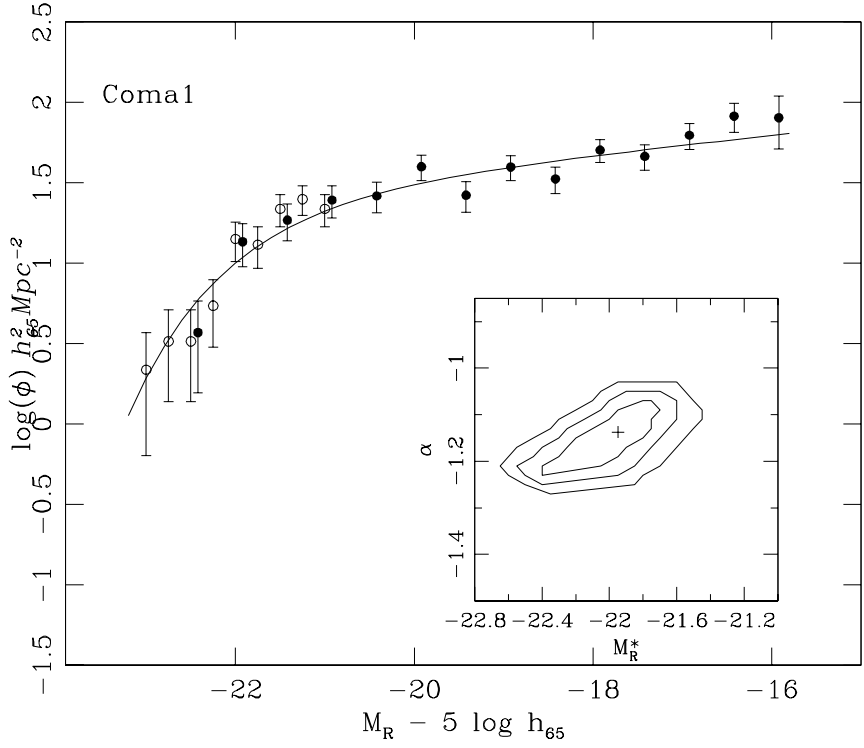}
\caption{Observed LFs for Coma1, Coma2 and Coma3 fields. The DSS+SSS
sample for a given field is shown by filled circles and the extended SSS
sample (normalized to DSS+SSS survey) by open circles. 
Schechter LF fits, as listed in Table 3, are also shown.
1,2 and 3$\sigma$ error contours for $M_R^\ast$ and $\alpha$ 
are presented on the separate panel.
\label{fig5}}
\end{figure}

\begin{figure}
\figurenum{5}
\plotone{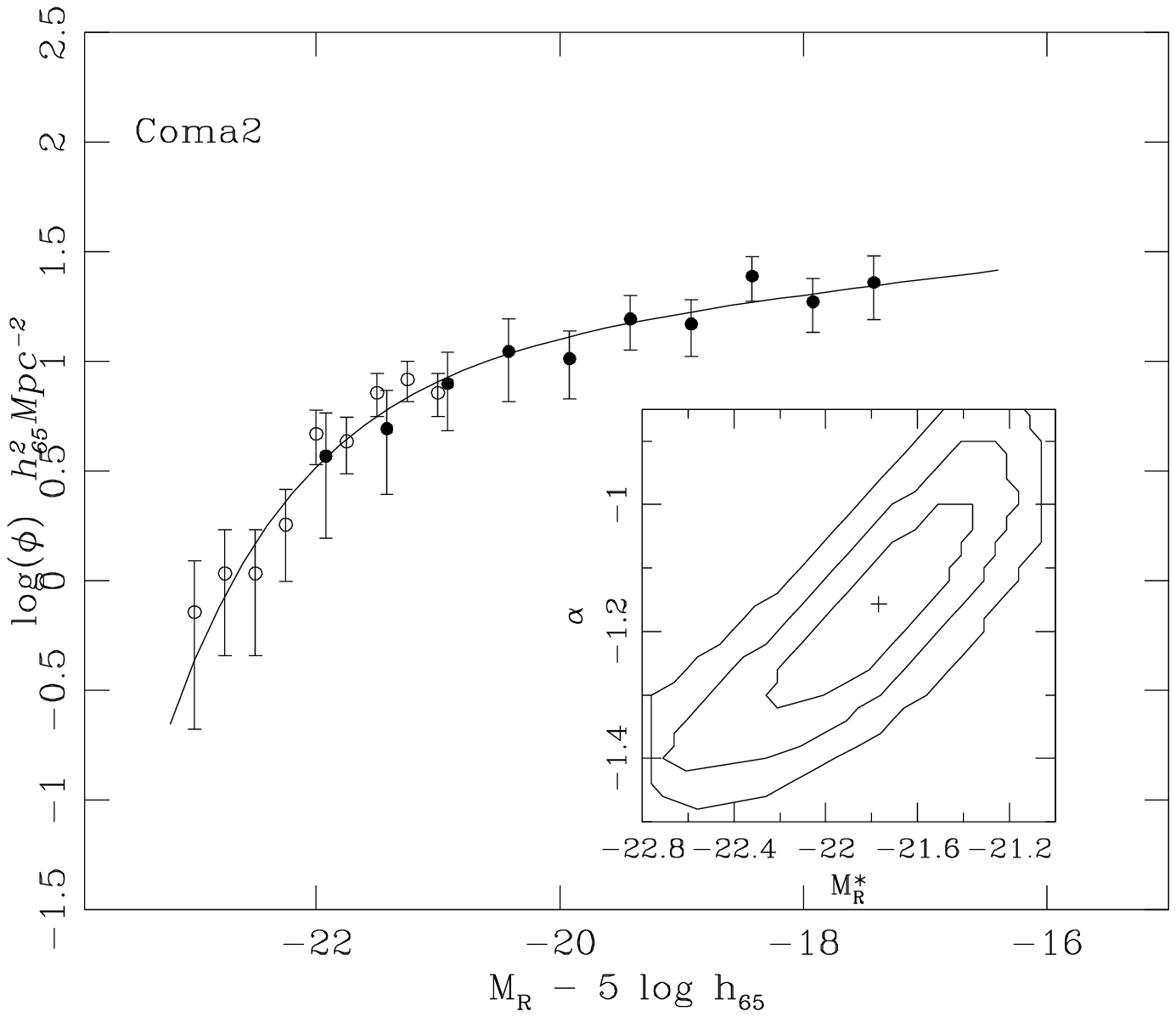}
\caption{}
\end{figure}

\begin{figure}
\figurenum{5}
\plotone{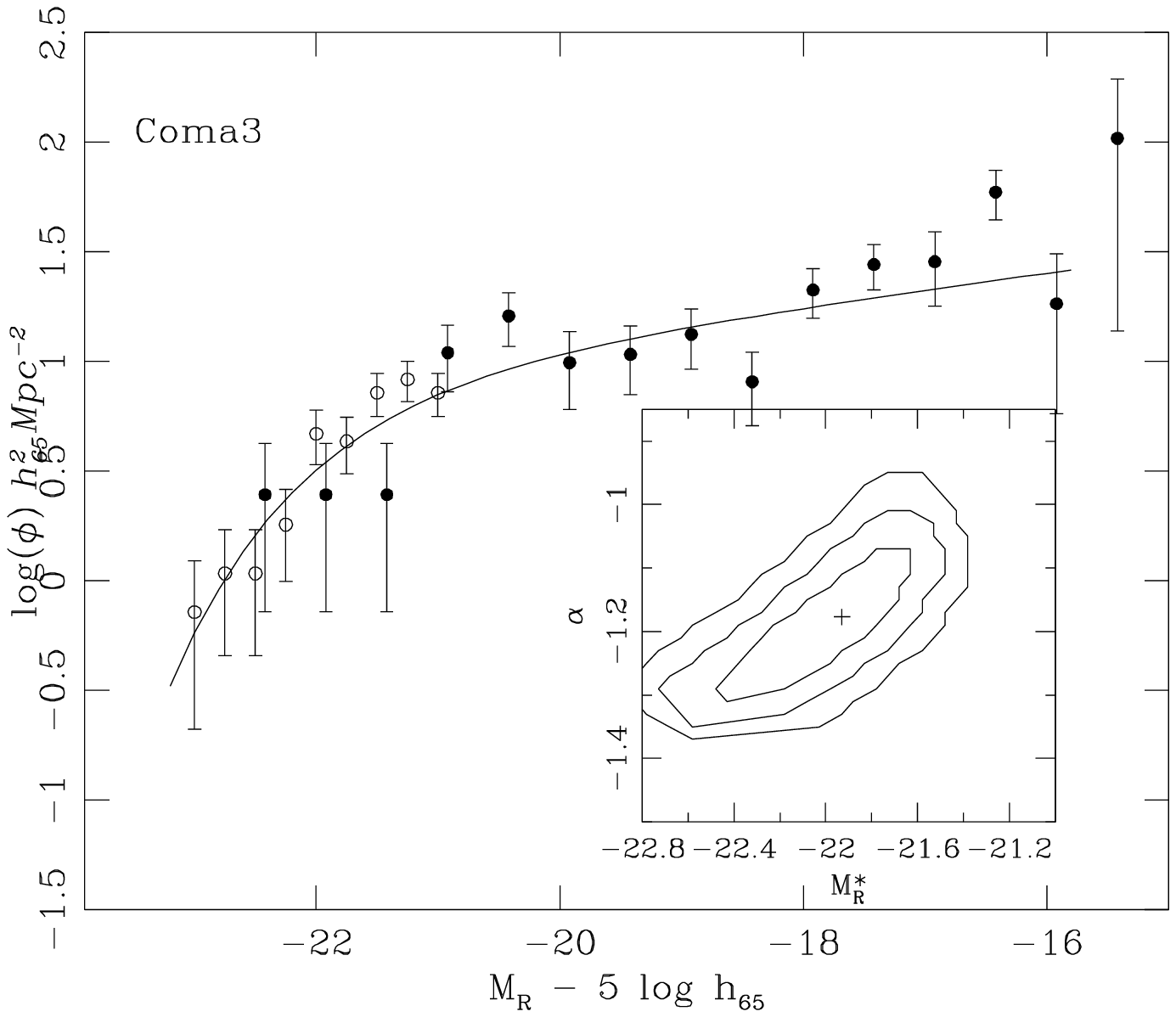}
\caption{}
\end{figure}

\subsection {Parametric fits to luminosity functions}

We consider a Schechter parametric form for the observed LFs, expressed as
$$\Phi_s (M)\ =\ \phi^*\ X^{\alpha + 1}\ e^{-X} $$

\noindent where $X=10^{-0.4(M-M^*)}$, with $M^*$ (characteristic 
magnitude), 
$\alpha$ (faint-end slope) and $\phi^*$ (normalisation) the parameters
to be determined (Schechter 1976). 

Galaxies from DSS and SSS, combined with those from 
the extended SSS sample, are used in the fit (both the solid and open circles
in Figure~5). As is clear from Table 3 and Figure 5, galaxies from the 
DSS+SSS and extended SSS samples overlap in the magnitude range
$-22.50 < M_R < -21.00 $. For galaxies over this interval, the average
of $\phi (M)$ values are calculated in each magnitude bin and used to fit
the LF. This allows the magnitude bins to be independent in the fitting
process. Using only one of the DSS+SSS or extended SSS samples over their
common magnitude interval, will not affect the results.  
The fits are carried out over the luminosity range 
$-23 < M_R - 5\ log\ h_{65} < -16$ for Coma1 and Coma3 and 
$-23 < M_R - 5\ log\ h_{65} < -17.5$ for
Coma2 fields. The fits for individual fields are shown in Figure~5 and
presented in Table 4. The reduced $\chi^2$ values, $\chi^2/\nu$, 
number of degrees of freedom, $\nu$, and the goodness of fit 
probability, $P(\chi^2|\nu)$, are also listed for each field. 
The 1,2 and 3$\sigma$ error contours corresponding to each of the LFs 
are shown in Figure~5, with the $1\sigma$ error estimates for LF parameters
presented in Table 4. The $\chi^2$ results in Table 4 indicate 
$P(\chi^2|\nu) > 20\%$ probability that the data, in all the three fields, 
are drawn from a Schechter LF, confirming that
a Schechter form provides acceptable fits to individual LFs 
both at the core and in the outskirts of the cluster. 

\begin{table*}
\caption{Results from the parametric fits to the observed LFs in Coma1, Coma2,
 Coma3 and the entire Coma cluster. $1\sigma$ error estimates are also 
listed. For each of the three fields, the LF parameters are estimated after
constraining their bright-end using data from the extended SSS sample 
(1st line for each field). Then
the $M_R^\ast$ values are fixed to these estimates and a 2-parameter fit
is performed to measure $\alpha$ and $\phi^\ast$ (2nd line for each field). 
$^\dagger$ Coma3 LF with the NGC4839 group removed.
\vspace{12pt}}
\begin{tabular}{cccrrrr}
        & $M_R^\ast - 5\ log\ h_{65}$ & $\alpha$ & $\phi^\ast$ (h$^2_{65}$\ Mpc$^{-2}$) & $\chi^2/\nu$ & $\nu$ 
& $P(\chi^2|\nu)$ \\
        &            &         &      &      &    &     \\
Coma1   &  $-22.00^{+0.13}_{-0.12}$  & $-1.15^{+0.03}_{-0.03}$ & $27.3^{+1.0}_{-1.5}$ & 0.92 & 16 & 65\% \\
        &            &         &      &      &    &      \\
Coma1   &  $-$       & $-1.17^{+0.03}_{-0.02}$ & $25.4^{+1.3}_{-1.0}$ & 0.93 & 12 & 65\% \\
        &            &         &      &      &    &      \\
        &            &         &      &      &    &      \\
Coma2   &  $-21.76^{+0.15}_{-0.12}$  & $-1.16^{+0.10}_{-0.10}$ & $11.9^{+1.0}_{-0.9}$ & 0.36 & 12 & 98\% \\
        &            &         &      &      &    &      \\
Coma2   &   $-$      & $-1.20^{+0.06}_{-0.04}$ & $10.5^{+1.0}_{-1.0}$ & 0.24 & 8 & 98\% \\
        &            &         &      &      &    &      \\
        &            &         &      &      &    &      \\
Coma3   &  $-21.98^{+0.15}_{-0.15}$  & $-1.19^{+0.05}_{-0.04}$ & $8.9^{+2.0}_{-1.7}$  & 1.29 & 17 & 20\% \\
        &            &         &      &      &    &      \\
Coma3   &  $-$       & $-1.29^{+0.04}_{-0.03}$ & $6.2^{+1.5}_{-1.5}$  & 1.63 & 13 & 10\% \\
        &            &         &      &      &    &      \\
Coma3$^\dagger$ & $-21.92^{+0.15}_{-0.15}$ & $-1.31^{+0.05}_{-0.05}$ & 
$3.1^{+1.2}_{-1.0} $ & 1.72 & 17 & 5\% \\
        &            &         &      &      &    &      \\
        &            &         &      &      &    &      \\
all fields & $-21.79^{+0.08}_{-0.09}$ & $-1.18^{+0.04}_{-0.02}$ & $9.5^{+0.5}_{-0.6}$ & 1.78 & 23 & 1\% \\
\end{tabular}
\end{table*}

The procedure of using the bright-end of the LF from the entire Coma cluster 
to constrain that of the individual fields (located at different positions
in Coma), makes the implicit
assumption that the LFs at bright magnitudes are independent of 
their local environment. To explore the extent to which this assumption 
could affect the overall shape of the LFs and their parametric form,  
we now fix $M^*$ values for each field to those measured and listed 
in Table 4, and carry out a two parameter fit to DSS+SSS (ie. the sample
which extends to fainter magnitudes) to determine $\alpha$ 
and $\phi^*$. These results together with their corresponding 
goodness of fit estimates are also presented in Table 4. There is no
difference (within the errors) in the faint-end slope and normalisation 
of LFs between the two cases, with both providing acceptable fits to 
the observed LFs. The probability that these data are drawn from a 
parent Schechter LF is $P(\chi^2|\nu) > 10\%$, further
confirming the previous result that a Schechter form provides
acceptable fit to the LF at the core and outskirts of the Coma cluster. 
This
justifies the above assumption that the bright-end of the LFs are similar
among the three fields in the Coma.

Using the techniques developed and the results so far, we now
concentrate on addressing the following three questions:

(i) Is the LF the same at the core and outskirts of the Coma ? how
does it depend on luminosity?

(ii) Can the LF for the entire Coma cluster be parametrised by a single form 
over its entire
luminosity range?

(iii) Does the LF change, depending on the observed wavelength or 
color and surface brightness of galaxies?

\subsection{Environmental dependence}

While the characteristic magnitudes for the LFs in the Coma1, Coma2 and Coma3
fields are close (Table~4), there is evidence for a slight 
increase in the faint-end slope
from $-1.16$ at the core (Coma1) to $-1.29$ at the outskirts (Coma3). 
However, this is
only a 1$\sigma$ effect. It is also clear from the $\chi^2$ tests that
a Schechter function form provides acceptable fits to the observed LFs in all 
three fields and over the entire magnitude range covered. Moreover, 
this shows that the three LFs have the same shape over their
common luminosity range. 

Given the similarity of LF shapes in different environments in the Coma 
cluster, it is appropriate to sum the individual LFs to estimate
the total LF, $\Phi (M)$, for the Coma cluster as

$$\Phi(M) = \Sigma_{i=1}^3 \phi_i(M) $$

\noindent While this provides an accurate estimate of the faint-end 
($M_R > -18$), it is only weakly constrained at the bright-end ($M_R < -21$), 
with large Poisson errors. Therefore, the total LF is derived here by
combining galaxies in the DSS+SSS samples (Coma1, Coma2, Coma3 fields), as
measured by $\Phi(M)$, with those in the extended SSS (all galaxies in the
shallow spectroscopic survey covering an area of $1^\circ$ diameter
in the Coma). The membership fraction is measured as discussed
in the last section. The two LFs are normalised over the magnitude range
$ -22.5 < M_R < -21 $, with only the extended SSS sample at 
$M_R < -17.75 $ used for fitting to the parametric form. 
The total LF is then derived 
over a magnitude range $-23 < M_R - 5\ log\ h_{65} < -16$ and
presented in Figure 6. 
The total LF over a 1-degree radius region about 
the cluster center is listed in Table 5, with its corresponding parametric 
fit presented in Table 4. 

The reduced $\chi^2$ value, corresponding to the LF from all fields
(last line in Table 4), indicates a small probability
($P(\chi^2|\nu)\sim 1\%$) that the data are drawn from a Schechter LF. 
We now consider the possibilities which might contribute to this. 
The total LF in Figure 6 is constructed assuming the LF
for galaxies covering the entire 1$^\circ$ diameter in the Coma to have
the same shape
as those in the three (Coma1, Coma2 and Coma3) fields. However, it is likely
that this is not the case for other regions of the Coma cluster (due to
differences in local densities), affecting the bright-end of the LF. 
Indeed, constraining the sample to only the three fields studied here 
(DSS+SSS galaxies located in Coma1, Coma2 and Coma3 fields), 
increases the probability to $P(\chi^2|\nu)\sim 10\%$, with no significant 
change in the estimated LF parameters listed in Table 4. Moreover, 
differences in normalisations ($\phi^\ast$ values) between the three fields
are likely to result in a poor fit to a Schechter LF form once the fields
are combined. Alternatively, 
it is possible that a 2-component LF (ie. Gaussian+Schechter) is needed
to model the total LF for the Coma cluster (Biviano et al 1995; 
Yagi et al. 2002). 

The observed faint-end slope of $-1.18^{+0.04}_{-0.02}$ found here, 
while consistent with individual fields (Coma1, Coma2, Coma3), 
is relatively shallower than
other studies of Coma cluster LF (Trentham 1998; Secker \&
Harris 1996; Beijersbergen et al.\ 2002). However, the spectroscopic LF here
has a much brighter magnitude limit ($M_R = -16$). Moreover, 
the measurements based on photometric surveys are mostly 
unconstrained at faint magnitudes due to small number statistics, 
incompleteness, uncertain background correction and contamination by
globular clusters (see section 5 for details). 

Study of the dependence of LFs on their local environments here 
could be hampered by the presence of the NGC 4839 group in the Coma3 field. 
This is examined by removing galaxies associated with this group 
from the Coma3 sample and then, measuring its LF. 
Defining NGC 4839 group as galaxies within a 
circle of radius 9 arcmin. centered on the cD galaxy NGC 4839 
(Beijersbergen et al. 2001), we construct the Coma3 LF after removing 
these objects
and then carry out a parametric fit to the Schechter LF form. 
The result is also
listed in Table~4 and shows a slight increase ($1\sigma$) in the faint-end 
slope of the Coma LF from the core to outskirts. Spectroscopic data on another
field in the outskirts of the Coma cluster at a similar distance from 
the center as Coma3 is needed to accurately measure changes in the 
faint-end slope with radius.

\begin{table*}
\caption{Total LF ($\Phi (M)$) for the entire Coma cluster. 
Galaxies from the SSS sample ($M_R < -17.5$; $M_B < -16.25$) 
are combined with 
data from DSS ($M_R > -17.5$; $M_B > -16.25$). 
Errors are estimated as discussed in the text. 
The last four lines correspond to the DSS survey.
\vspace{12pt}}
\begin{tabular}{cccrcc}
$M_R$     & $\Phi^R (M)$ & $\Delta (\Phi^R (M))$ & $M_B$ & $\Phi^B (M)$ & 
$\Delta (\Phi^B (M))$ \\
       &      &      \\
\multicolumn{6}{c}{Extended SSS} \\
       &      &      \\
-23.00 & 0.72 & 0.51 & -21.75  &  0.72  &   0.51 \\
-22.75 & 1.08 & 0.62 & -21.25  &  1.44 &  0.72 \\
-22.50 & 0.36 & 0.36 & -21.00  & 0.72  &  0.51 \\
-22.25 & 1.80 & 0.81 & -20.75  &  2.16 &  0.88 \\
-22.00 & 4.69 & 1.30 & -20.50  &  2.52 &  0.95 \\
-21.75 & 4.33 & 1.25 & -20.25  &  3.96 &   1.19 \\
-21.50 & 6.13 & 1.49 & -20.00  &  5.77 &   1.44 \\
-21.25 & 8.29 & 1.73 & -19.75  &  9.01 &   1.80 \\
-21.00 & 6.85 & 1.57 & -19.50  &  7.57 &   1.65 \\
-20.75 & 11.17 & 2.01 & -19.25 &   10.81 &    1.97 \\
-20.50 & 10.81 & 1.97 & -19.00 &   12.61 &   2.13 \\
-20.25 & 12.61 & 2.13 & -18.75 &   15.14 &   2.33 \\
-20.00 & 11.17 & 2.01 & -18.50 &   11.17 &   2.01 \\
-19.75 & 8.65 & 1.77  & -18.25 &  10.09  &  1.91 \\
-19.50 & 10.81 & 1.97 & -18.00 &   9.37  &  1.84 \\
-19.25 & 9.73 & 1.87 & -17.75  &  11.89  &  2.07 \\
-19.00 & 13.95 & 2.25 & -17.50 &   17.21 &   2.51 \\
-18.75 & 13.21 & 2.19 & -17.25 &   15.19 &   2.36 \\
-18.50 & 17.57 & 2.54 & -17.00 &   14.84 &   2.35 \\
-18.25 & 17.29 & 2.52 & -16.75 &   13.88 &   2.28 \\
-18.00 & 12.30 & 2.19 & -16.50 &   16.25 &   2.51 \\
-17.75 & 16.80 & 2.68 & -16.25 &   12.28 &   2.26 \\
       &      &      \\
\multicolumn{6}{c}{DSS Survey} \\
       &      &      \\
-17.50 & 21.23 & 1.75 & -16.00 & 12.28 &   1.14\\
-17.00 & 19.96 & 2.11 & -15.50 & 13.31 &   1.43\\
-16.50 & 31.00 & 3.06 & -15.00 & 18.93 &   1.97\\
-16.00 & 21.65 & 4.30 & -14.50 & 11.62 &   2.13\\
\end{tabular}
\end{table*}

\begin{figure}
\figurenum{6}
\plotone{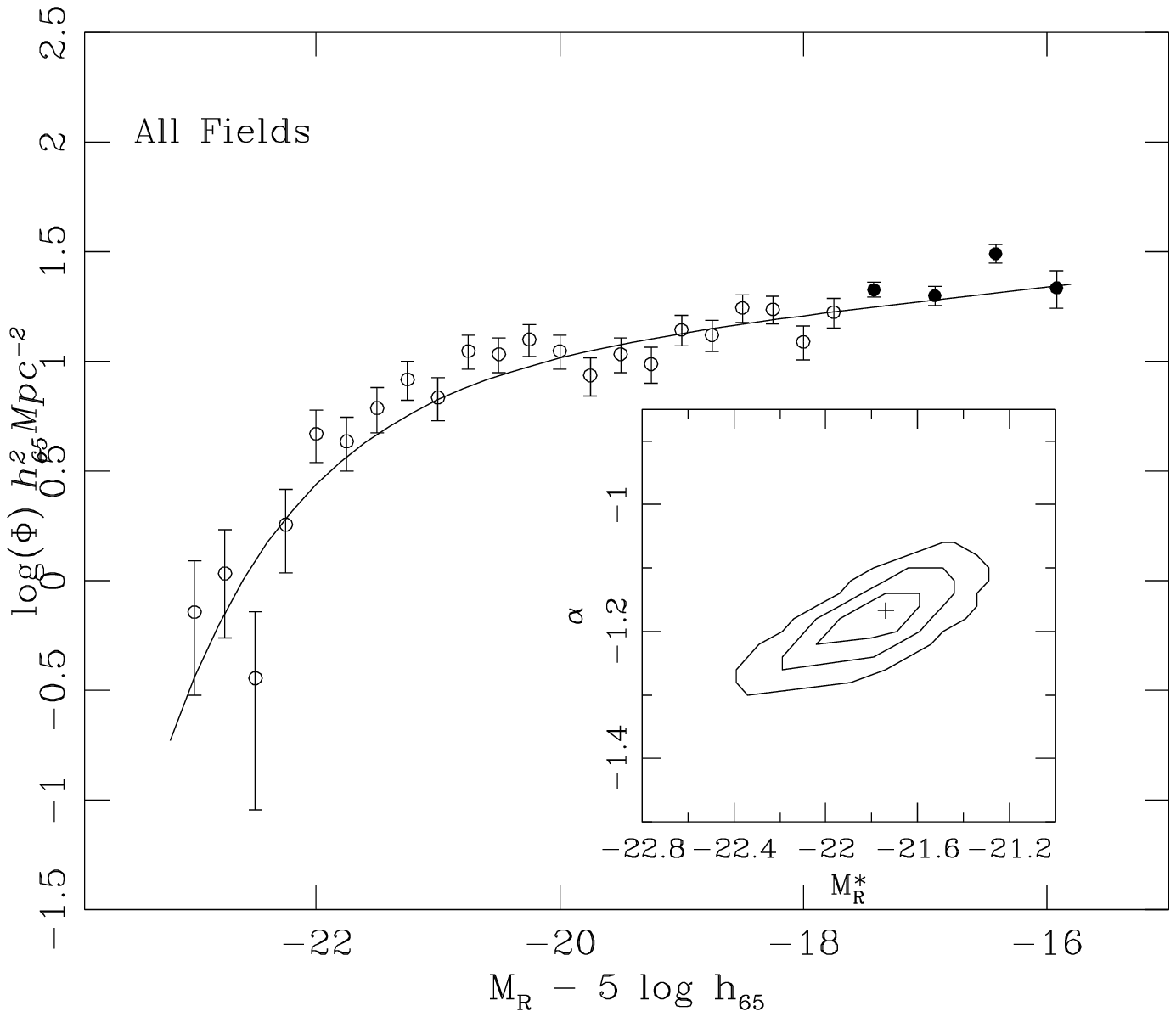}
\caption{Total LF, covering an area of 1$^\circ$
radius of the Coma cluster. Combined sample from the SSS (open circles)
and DSS (filled circles) are used. Schechter LF fit is also presented.
$M_R^\ast$ and $\alpha$ error countours at 1,2 and 3$\sigma$ levels 
are also shown.}
\label {fig6}
\end{figure}

\subsection{Wavelength Dependence}

The B-band LF for the entire Coma cluster is constructed 
and presented in Figure 7, following 
the same procedure discussed in section 4. The extended SSS (covering
1 deg$^2$ area of the Coma) and DSS samples are used with the completness
functions estimated in the same way as for the R-band LF. The B-band
magnitudes which were not available for a fraction of galaxies in 
the extended SSS survey are estimated using their R-band
magnitudes (measured from photographic plate $b$ and $r-$band 
data in section 4) and $B-R$ colors. The total B-band LF is also tabulated 
in Table 5.

The spectroscopic sample in this study was selected in R-band. This was
motivated because the light at longer wavelengths is dominated by 
evolved stellar population and because the Coma cluster is rich in
early-type evolved galaxies. Therefore, we expect a bias against the
very blue and faint galaxies, leading to a flatter  faint-end slope
for the B-band LF here. Having this in mind, the B-band LF is fitted
to a Schechter form, with the best fit and 1,2 and 3$\sigma$ error contours
presented in Figure 7. We estimate $M_B^\ast = -19.95^{+0.15}_{-0.10} 
+ 5\ log\ h_{65}$; 
$\alpha = -0.96^{+0.01}_{-0.02}$;
$\phi^\ast=16.3^{+0.1}_{-0.1}\ h_{65}^2\ Mpc^{-2}$ 
and $\chi^2/\nu = 1.16 $. This corresponds
to $P(\chi^2 | \nu) = 25\%$ probability that a Schechter LF form is a 
good representation of the data.

The total B-band LF in Figure 7 shows a dip at $M_B = -18$ mag. A similar gap
with smaller amplitude has been observed in the total R-band LF (Figure 6) 
at almost
the same magnitude, $M^\ast_R= -19.5$, shifted by the mean color of galaxies
in this sample ($<B-R>=1.5$). Although this is only a 1$\sigma$ effect and
likely caused by Poisson statistics, a detailed study of this feature
is useful as it gives clues towards the 
overall shape of the LF. Such a behavior has also been 
confirmed from other studies at almost the same magnitude in the rest-frame 
B-band LFs of both nearby (Biviano et al 1995) and 
intermediate redshift (Dahlen et al 2003) clusters. Studying the R-band LFs
for a photometric sample of 10 clusters at different redshifts and 
with different richness classes, 
Yagi et al (2002) show evidence for a dip in the cluster LFs 
at the same magnitude as here. After dividing their sample into type specific
groups, consisting of elliptical ($r^{1/4}$-like profile) and spiral
(exponential-like profile), they conclude that the observed dip is almost 
entirely due to contribution from the early-type ($r^{1/4}$-like) galaxies
to the total LF. They also find that clusters with larger velocity dispersions
have more distinct dips, perhaps indicating that the amplitude of 
the dip depends on the dynamical state of the cluster and consequently, on
the dominant population of galaxies in that cluster. 

To explore the effect of this gap on the general shape of the LF, 
we now remove galaxies in the range $-18.5 < M_B < -17.75$, which contribute
to the gap in Figure 7 and then fit the LF to a Schechter form. We find
$M_B^\ast = -19.74 \pm 0.10 + 5\ log\ h_{65}$; 
$\alpha = -0.90 \pm 0.02$;
$\phi^\ast=20.7\pm 0.1\ h^2\ Mpc^{-2}$ and $\chi^2/\nu = 0.97 $. 
This increases the
probability of a Schechter LF form being a 
good representation of the data to $P(\chi^2;\nu) = 50\%$. Given this
result, it is likely that
a two-component shape for the LF, consisting of a Gaussian, representing
giant galaxies ($M_B < -18$), and a power-law, 
representing the dwarf population ($M_B > -18$), may well be appropriate
for the composite LF over its entire magnitude range.

\begin{figure}
\figurenum{7}
\plotone{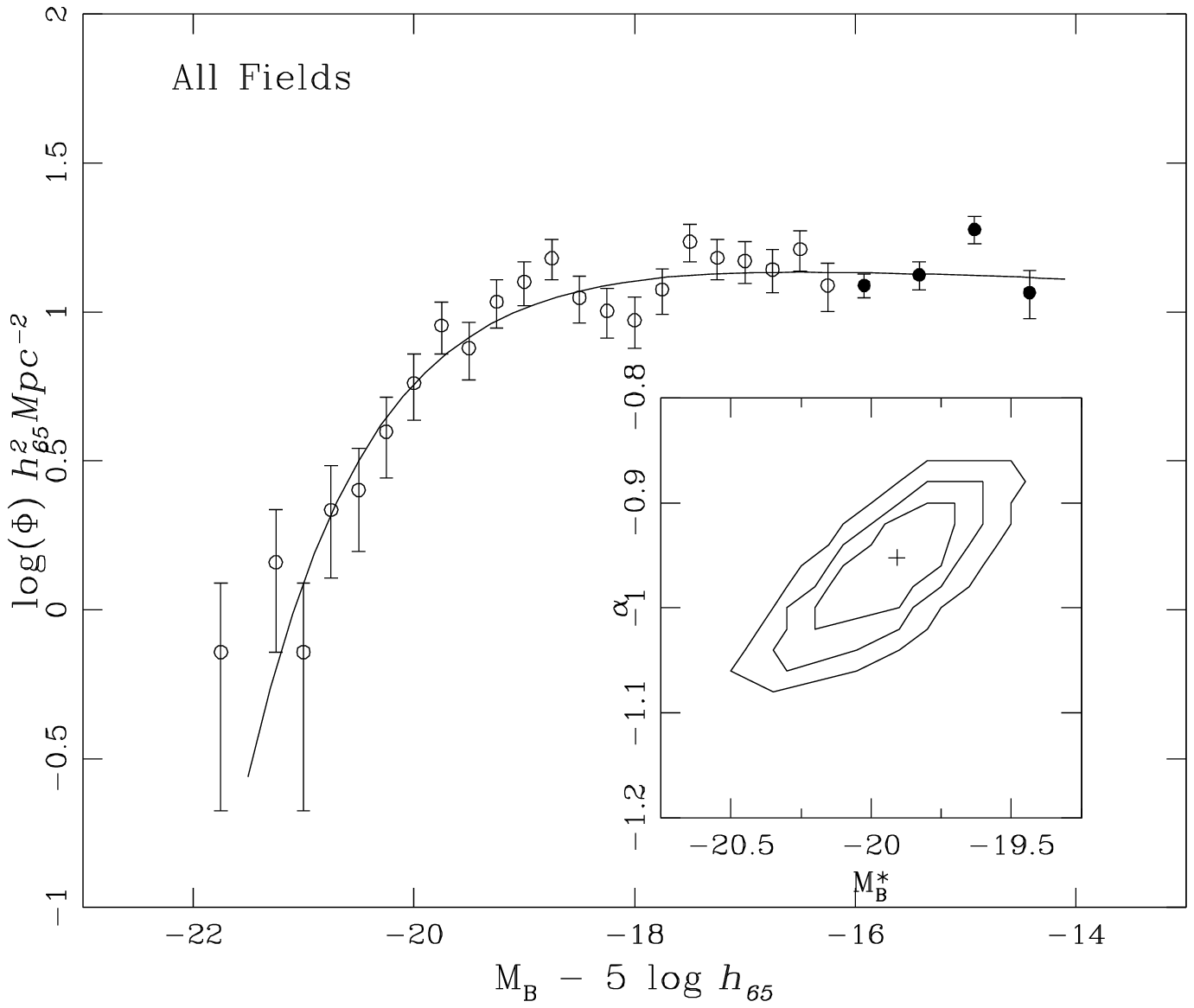}
\caption{Total B-band LF, covering an area of 1$^\circ$
radius of the Coma cluster. Combined sample from the SSS (open circles)
and DSS (filled circles) are used. Schechter LF fit is also presented.
$M_R^\ast$ and $\alpha$ error countours at 1,2 and 3$\sigma$ levels 
are also shown.}
\label {fig7}
\end{figure}

\subsection{Color Dependence}

The survey performed here is extensive enough to allow a study
of the LFs both in different environments in the Coma cluster and 
in color intervals. This requires knowledge of the membership fraction in
$B-R$ color intervals. To minimise statistical uncertainties, we use
galaxies from all the fields (in both DSS and SSS samples) to derive a common
selection function, assuming no spatially-dependent selection bias.  

\begin{figure}
\figurenum{8}
\plotone{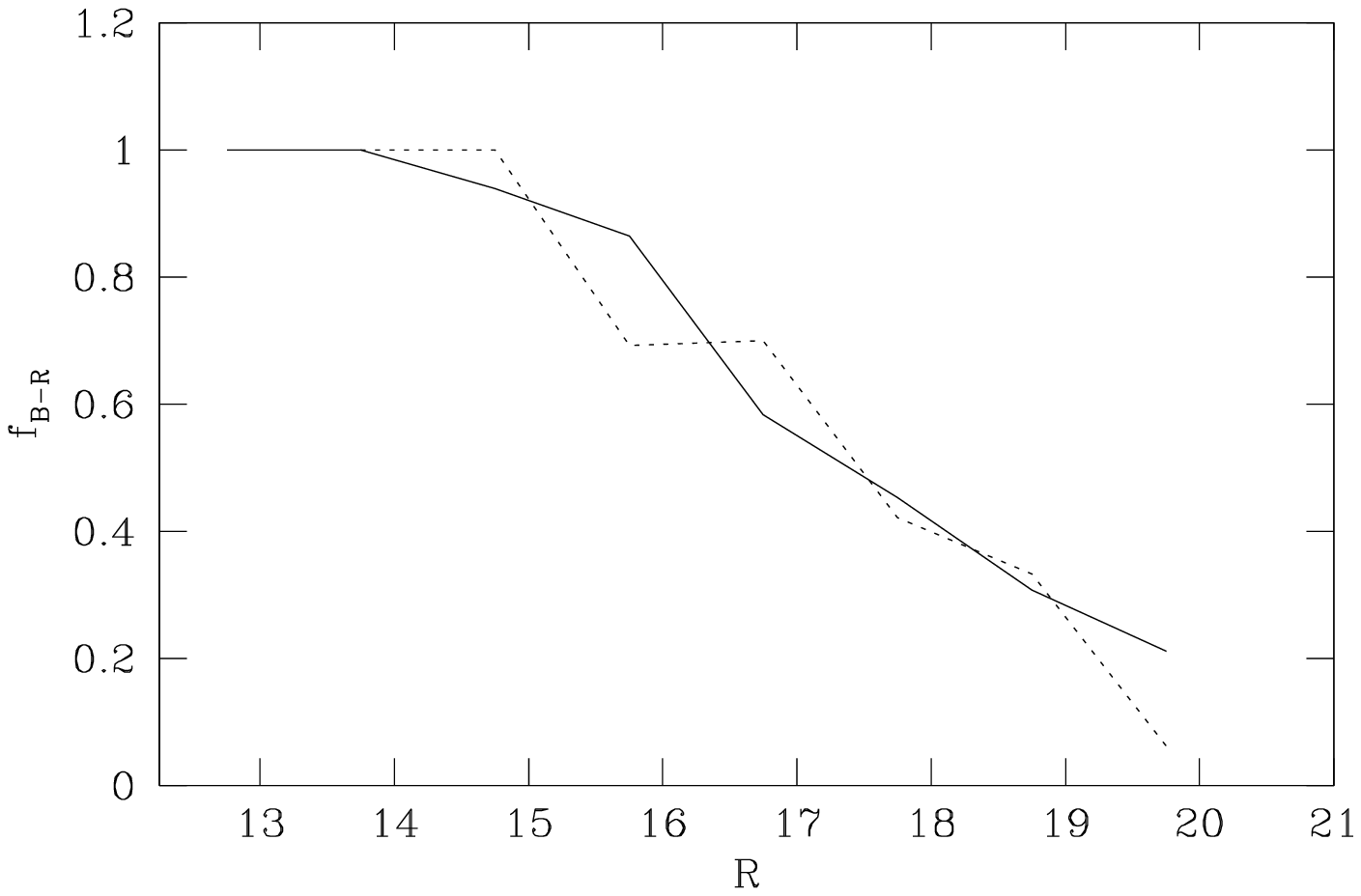}
\caption{Membership fractions in $B-R$ color intervals
for galaxies in DSS+SSS samples; $B-R > 1.35$ (solid line) and
$B-R < 1.35$ (dashed line).}
\label {fig8}
\end{figure}

Two color intervals are adopted, corresponding to blue ($B-R < 1.35$) and
red ($B-R > 1.35$) galaxies. This color is chosen so that it is
consistent with the average color of an intermediate type spiral (ie. Sc), 
allowing for clear separation between star-forming (blue) and evolved (red)
populations. Moreover, this provides sufficiently large number of
blue and red galaxies for statistical analysis. 
The blue galaxies here are also 
found to have strong emission lines, further confirming that they
are undergoing star formation activity. The membership fractions, 
derived for the two color bins following the 
procedure explained in section 3, are
presented in Figure 8. It is clear that, for both blue and red galaxies, 
the survey is 80\% complete to $R\sim 16$, droping to 40\% completeness at 
$R\sim 18.5$. Using membership fractions, the LFs are derived in
$B-R < 1.35$ (star-forming) and $B-R > 1.35$ (evolved) intervals and
presented in Figure 9 for the three fields separately
and all the fields together. 

\begin{figure}
\figurenum{9}
\plotone{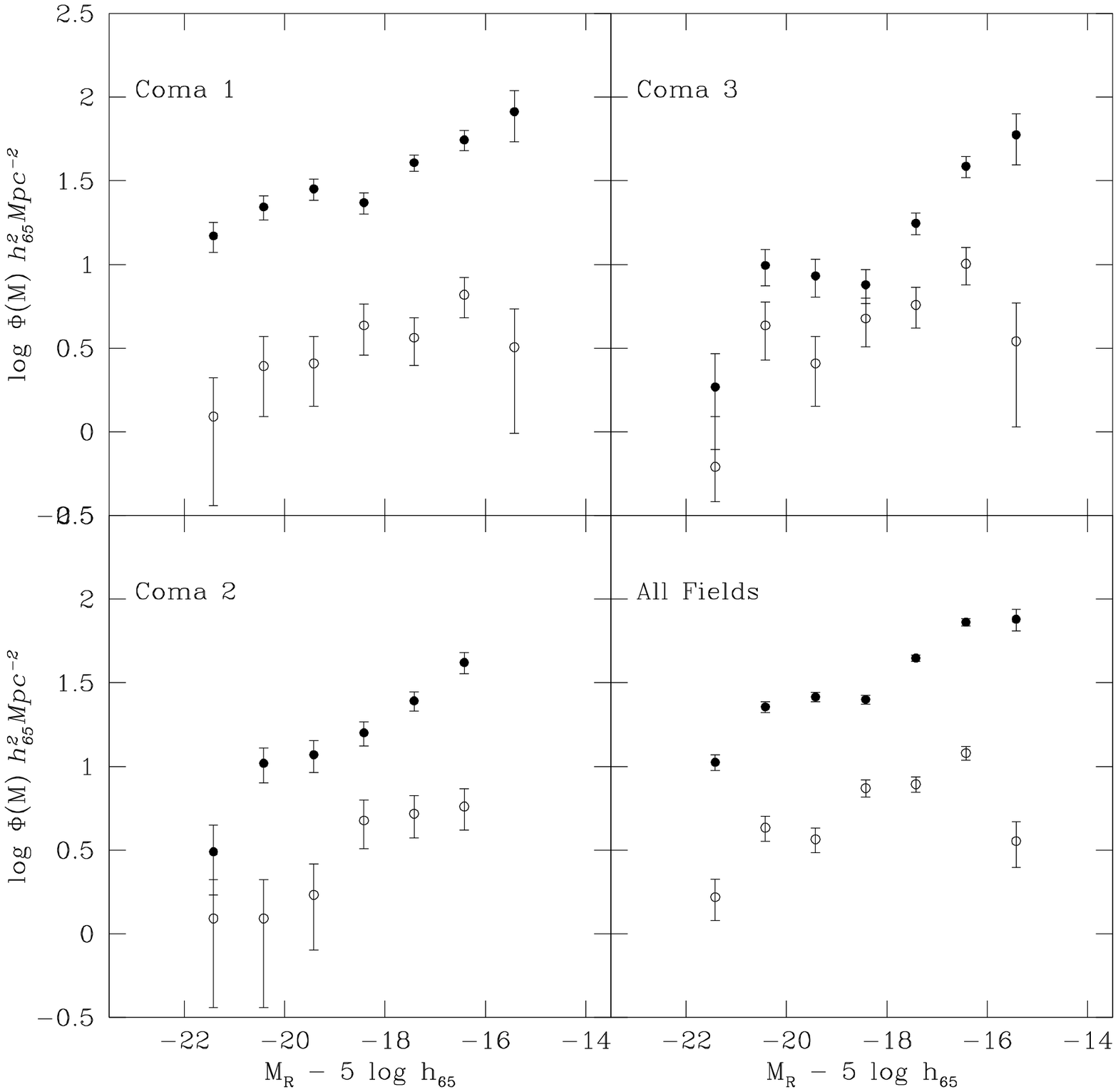}
\caption{LFs in $B-R$ color intervals for galaxies at the core
and outskirts of the Coma cluster. $B-R > 1.35$ (filled circles) and
$B-R < 1.35$ (open circles). The errorbars are estimated as discussed 
in the text.}
\label {fig9}
\end{figure}

The evolved galaxies in Coma1 have a surface density which is an order of
magnitude higher than that of star-forming galaxies, with the difference
decreasing towards the outskirts (Coma2 and Coma3 fields). 
The sudden increase 
in the slope of the Coma3 field LF for red galaxies at $M_R > -18$ is due to 
the contribution from the NGC4839 group. The $B-R$ color distribution for 
spectroscopically confirmed members of this group (Figure 10) shows a
significant number of these galaxies having $B-R > 1.35$.  
The similarity
between the slope of the LFs for star-forming galaxies in different 
environments contradicts the result from Beijersbergen et al.\ (2002) 
who found a relatively steeper U-band LF in the outerpart of the Coma cluster.
The difference here is likely due to selection of the present sample in the
red passband, statistical fluctuations in both studies and uncertainties
in background subtraction in the Beijersbergen et al. sample. 

Recently, it was shown that for a given spectral type of galaxies, the
faint-end slope of the cluster LF is steeper than that in the field
(De Propris et al 2003). This is in apparent contradiction with the
result here, in which similar LFs are found for both the blue and red
populations in the core (Coma1) and outskirts (Coma3) of the Coma. However, 
clusters used in De Propris et al (2003), cover a wide range in richness, 
with many being as rich as the Coma3 field here. This indicates a larger
density contrast between the cluster sample of De Propris et al and the
general field, compared to the Coma1 and Coma3 fields. Moreover, the
results from the two studies are not directly comparable as there is only
a loose relation between colors and spectral types used in these two studies.

\begin{figure}
\figurenum{10}
\plotone{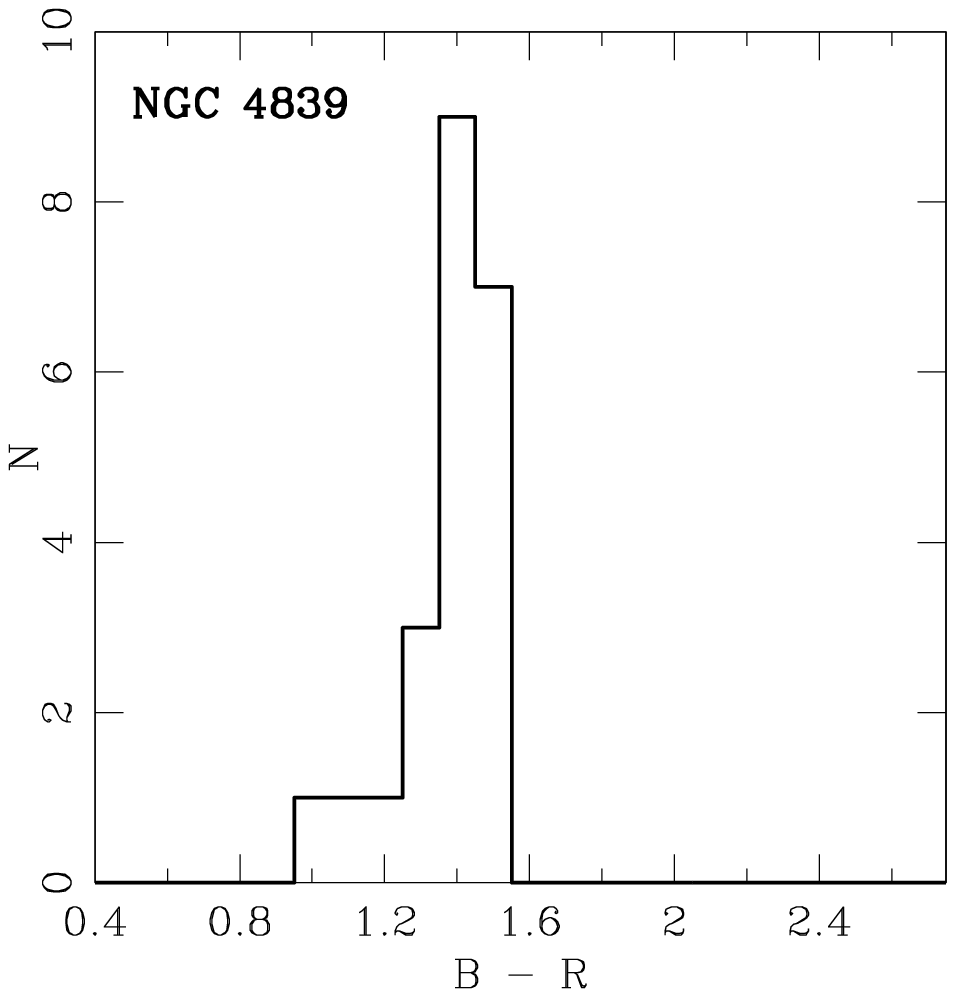}
\caption{$B-R$ color distribution for spectroscopically confirmed members
of the NGC4839 group in the Coma3 field.}
\label {fig10}
\end{figure}

\subsection{Surface Brightness Dependence}

The surface brightness distributions 
in Figure 3 show that, on average, the cluster members have a relatively
brighter surface brightness compared to field galaxies. 
This implies that higher surface brightness galaxies tend to reside in
environments with relatively larger densities. 
Figure 11 presents luminosity distributions for spectroscopically
confirmed members of the Coma cluster in three 
surface brightness intervals; $\mu_{eff} < 20$, $20 < \mu_{eff} < 22$ and
$\mu_{eff} > 22$ mag./arcsec$^2$. There is a clear trend in the sense
that low surface brightness galaxies have fainter magnitudes, reflecting
a monotonic magnitude-surface brightness relation. Such relation has also
been found for field galaxies (Brown et al 2001; Cross et al 2001). 
However, there is also a considerable 
number of low surface brightness ($\mu_{eff} > 22$) galaxies in all
the Coma fields ($> 50\%$)- (Figure 11). This
indicates a low surface brightness population that 
dominates the faint-end of the LF, in
agreement with results from Andreon \& Cuillandre (2001) and
Sprayberry et al (1997). 

\begin{figure}
\figurenum{11}
\plotone{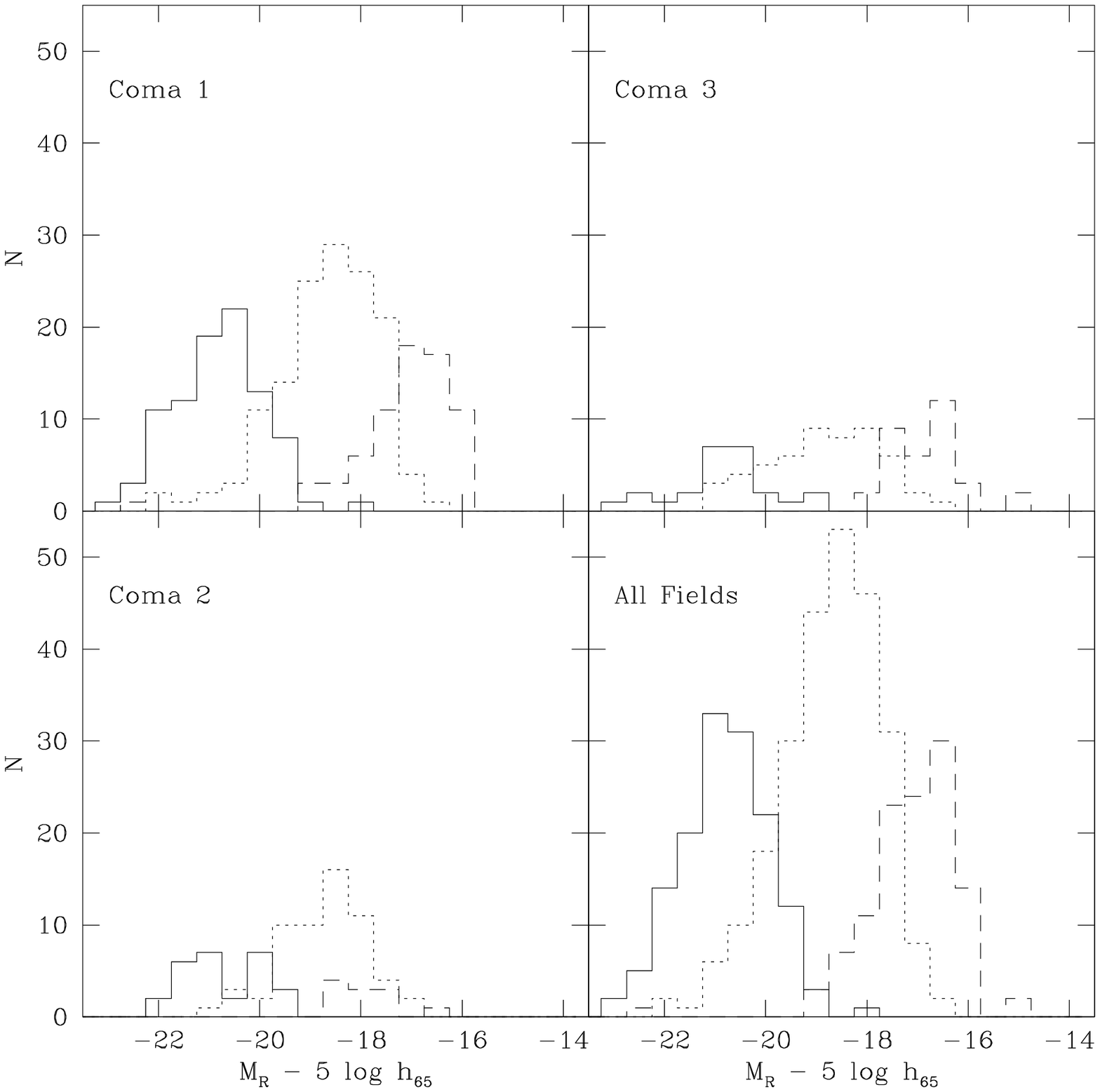}
\caption{Absolute magnitude distributions in effective surface brightness 
intervals for galaxies at the core, 
outskirts and the entire Coma cluster; 
$\mu_{eff} < 20$ (solid line), $20 < \mu_{eff} < 22$ (dotted line)
and $\mu_{eff} > 22$ (dashed line) mag/arcsec$^2$.
}
\label {fig11}
\end{figure}

\section {Comparison with other Studies}

Recent studies of the Coma cluster LF have led to steep faint-end
slopes in both optical (Thompson and Gregory 1993; Secker \& Harris 1996; 
Trentham 1998) 
and near-IR (Mobasher \& Trentham 1998; De Propris et al.\ 1998; 
Rines et al.\ 2001) wavelengths. 
All these studies are based on photometric surveys
with statistical background corrections from control fields and varying 
fields of view. 
Recently, this technique has gained popularity by the
advent of large-format and highly sensitive panoramic detectors, 
allowing more statistically 
representative measure of background contamination. 

Using photographic plates, Thompson and Gregory (1993) determined 
the LF to $B\sim 20$ ($M_B\sim -15$). They found a steepening of the slope
at faint magnitudes ($\alpha =-1.4$). This was followed by Bernstein et al.\
(1995) study, who used CCD surveys, extending the Coma LF to much fainter 
limits ($M_R \sim -9.5$)
over a smaller region. These authors also found a steep faint-end slope
($\alpha= -1.8$). However, their result was significantly affected by 
globular cluster contamination in the halo of NGC4874, making their LF
unconstrained fainter than $M_R=-14$. A deeper survey over a much larger
area (700 arcmin$^2$) by Secker \& Harris (1996) also found a LF that
rises to $M_R=-14$ and flattens thereafter. The flat bright-end, followed
by steep faint-end slope was further confirmed by Lobo et al.\ (1997). 

Recently, Beijersbergen et al.\ (2002) carried out a multi-wavelength (UB$r$)
study of LF in the Coma cluster. They found $M_R = -21.47^{+0.12}_{-0.17}
\ + 5\ log\ h_{65}$ 
and 
$\alpha=-1.16$, with a marginally steeper faint-end slope 
at larger radii from the cluster
center. The increase in the faint-end slope of U-band LF was interpreted as
due to a population of star-forming dwarf galaxies. No such increase in
the number density of this population is found, as is clear from 
the B-band LF in Figure 7.  Also, 
Andreon and Cuillandre (2001) carried out a wide-area (650 arcmin$^2$) 
photometric survey
of the Coma cluster. Combined with HST observations, they confirmed that 
unresolved globular clusters could be mistakenly classified as dwarf galaxies
at faint magnitudes, causing the observed steepness of the LF faint-end 
slope. 
Correcting for this, they derived $\alpha =-1.4$, with both the shape and 
slope of the LF depending on color.
They also derived LFs in surface brightness intervals and found low 
surface brightness galaxies as the main contributor at fainter magnitudes, 
as expected from, for example, studies of the local group (Mateo 1998).   
This result is in close agreement with that found in section 4.5,   
using a spectroscopic sample, and that in Trentham (1998) from a photometric
survey.  

Trentham (1998) avoided some of the problems affecting the above studies 
by using a deeper control field to improve background estimates 
and contamination (at the faint-end) by globular clusters.  
Surveying a large area (674 arcmin$^2$) to deep ($R=24$) limits,  
he found a steep faint-end slope ($\alpha = -1.7$) 
at $R > - 17.6$ followed by a turn over in the LF at $R=-15.6$, in both
the core ($r < 200$ kpc) and outskirt ($r > 200$ kpc) fields. 
The population
of galaxies at $R\sim 19$ were found to be dwarf spheroidals with
$B-R = 1.3$. The narrow spread in $B-R$ colors here was taken as
evidence for a homogeneous population of dwarfs in Coma, in agreement
with recent results. 
However, the slope and general shape of the LF at faint magnitudes only
depend on a few points with large errorbars, making this part of the LF 
essentially unconstrained. 

All the above studies are based on photometric surveys with statistical
background corrections. The faint-end slopes of the LFs from photometric
surveys discussed above, are relatively steeper than that found from the
present study, based on a spectroscopic survey. This is mostly due to
the depth of the photometric surveys, extending to $M_R\sim -12$, 
as compared to the 
shallower spectroscopic survey here ($M_R < -16$), with the up-turn in the
faint-end slope appearing at $M_R > -16$. Moreover, it
is possible that a combination of uncertain
background correction and contamination by globular clusters, 
specially at faint magnitudes, may have led
to the steep faint-end slopes found in studies based on photometric surveys. 
However, in a recent study, using deep simulated catalogues of cluster
galaxies constructed from a flat LF ($\alpha=-1$), Valotto et al (2001) 
show strong tendency towards steep faint-end slope for LFs ($\alpha > -1.5$)
in clusters selected in two dimensions. This indicates the cause of the 
observed steep faint-end slope in clusters, seen in photometric surveys, as
due to projection effect. 

Using a sample of 205 spectroscopically confirmed galaxies with
$b_{26.5} < 18$ at the 
$48\times 25$ arcmin$^2$ center of the Coma cluster, Biviano et al (1995) 
found 
that a single Schechter function cannot adequately fit the LF of the Coma
cluster. He found that a combination of Schechter and Gaussian forms 
provides a significantly better fit to the LF. However, Biviano et al's data
do not strongly constrain the faint-end of the LF, with his spectroscopic
sample becoming progressively incomplete towards fainter magnitudes.

It is useful to compare the Coma LF derived here, with those in other
rich clusters (Virgo), less evolved clusters (Ursa Major) and the general
field. A recent study of the Virgo LF finds a steep faint-end slope of
$\alpha = -1.6$, based on photometric B-band CCD survey 
(Trentham \& Hodgkin 2002). Although steep, this is significantly shallower 
than the slope of $\alpha =-2.2$, found in $R$-band by Phillipps et al.\ 
(1998). Despite the fact that these are based on power-law fits 
over restricted magnitude range, 
both the studies show the LF becoming relatively flat after the rise
at the bright-end and before steepening towards fainter magnitudes. 
However, due to dominance of the star over galaxy counts and steep galaxy
counts in the Virgo, measurement of the LF for this cluster is uncertain. 
Therefore, the difference in the faint-end slopes found between these studies
shows the
uncertainty involved in measuring background contamination when using
photometric surveys in clusters (the optimal redshift for background 
subtraction is found to be $z\sim 0.15$ (Driver et al 1995)), different 
selection effects and methodologies. Moreover, an accurate determination of
the Virgo LF is further complicated by the
prolate shape of this cluster (Yasuda et al 1997; West \& Blakeslee 2000; 
Arnaboldi et al 2002).  
The Ursa Major LF is found to be significantly different from those in
richer clusters like Virgo and Coma (Trentham et al.\ 2001). For example, 
it does not show the 
steep increase at the faint-end, indicating a relatively smaller population
of dwarfs with respect to giants. 

Recently, there have been two comprehensive studies of the field LF from
SDSS (Blanton et al 2001) and 2dFGRS (Cross et al 2001; Norberg et al 2002). 
Because of
the need for spectroscopy in estimating the LF for field galaxies, these 
normally span a brighter magnitude range than their cluster counterparts, 
derived from photometric surveys with statistical background correction. 
This, combined with the fact that these two studies are based on different
filters and magnitude scales compared to present work, makes a detailed
and quantitative comparison difficult. Nevertheless, the faint-end slope 
of the Coma LF from this study ($-1.18^{+0.04}_{-0.02}$) agrees with  
those from the SDSS ($-1.24\pm 0.05$) and 2dFGRS ($-1.21\pm 0.03$) field LFs. 
Moreover, the SDSS reveals a turn-over in the faint-end of the LF 
for red and a relatively steep LF for blue galaxies. This effect is not
observed in Figure 9. However, in both the SDSS and 2dFGRS, strong
correlations are found between luminosity and surface brightness, in
qualitative agreement with results from the present study.

\section {Discussion}

The shape of the LF at faint magnitudes depends on the efficiency with
which gas is converted into stars in low mass systems (Dekel \& Silk 1986; 
Efstathiou 2000). 
For example, dark halos in dense
environments could collect gas (through intracluster pressure) 
and be turned into stars but dark halos in
unevolved (diffuse) environments could not, leading to more low luminosity
galaxies (dwarfs) per unit total mass in evolved compared to 
unevolved or field environments (Tully et al 2001; Somerville 2001). 
However, this effect is contrasted 
by the on-going process of galaxy interaction and tidal stripping in 
dense regions of clusters, resulting in a decrease in the number of
faint galaxies and hence, flatter faint-end slope for core LFs,  
while the reverse is the case in areas with small galaxy densities
(poor clusters and groups)- (Lopez-Cruz et al. 1997; Phillipps et al 1998; 
Adami et al. 2000; Conselice 2001). 
The observed faint-end slope of the LF is likely to be fixed by a 
competition between these two effects. 

The R-band LFs measured here show a slight increase in their faint-end slope
from $-1.16^{+0.03}_{-0.02}$ in Coma1 (inner region) to 
$-1.21^{+0.06}_{-0.04}$ in Coma2 (intermediate region) 
and $-1.29^{+0.04}_{-0.03}$ in Coma3 (outer region). However, this is only
a 1$\sigma$ effect. Contribution from the NGC4839 group to the faint-end
slope of the Coma3 field is examined by removing galaxies associated with
this group and estimating the Coma3 LF again. This does not significantly 
change 
the faint-end slope of the Coma3 field. Recently, using a spectroscopic 
sample of 60 clusters with different richness classes, selected from the 
2dFGRS, 
De Propris et al (2003) found a composite LF with  $\alpha=-1.25\pm 0.03$, 
similar to that of the field LF. However, for a given spectral type of 
galaxies, they found the cluster LF to be significantly steeper than the LF
for field galaxies of the same spectral type. 
The R-band luminosity of galaxies, 
used in the present study, is sensitive to the old stellar population (ie. mass
function) and hence, not significantly affected by on-going star formation. 
Therefore, the similarity of the LFs between the core and outskirts of the
cluster, and their closeness to the field LFs, estimated from SDSS 
(Blanthon et al 2001) and 2dFGRS (Cross et al 2001; Norberg et al 2002),  
suggests that the average luminosity produced for a given galaxy mass is
little affected by the processes which make significant changes to the
morphologies and spectral types of individual galaxies. This is in
qualitative agreement with results in De Propris et al (2003).

The similarity of the faint-end slopes of the R-band LF between core and 
outskirts of the Coma cluster is in agreement with the results 
in Trentham (1998)
who uses a significantly deeper photometric survey, extending to $M_R=-12$. 
However, Beijersbergen et al (2002) found an increase in the faint-end slope
of U-band LF with radius from the center of the Coma cluster, indicating an 
abundance of star-forming galaxies in low density regions. Considering 
the observed LFs from these studies, 
the theoretical prediction of a turn over in the LF at faint magnitudes, 
due to the suppression of star formation in low mass galaxies 
(Efstathiou 1992; Chiba \& Nath 1994; Thoul \& Weinberg 1995) is not 
supported. However, due to the limited depth of the spectroscopic survey
in this paper ($M_R < -16$) and uncertainties in background correction
and normalisation 
in photometric surveys (Trentham 1998; Beijersbergen et al 2002), one cannot
yet constrain these models. 
A single component Schechter function form gives acceptable fits to the
observed LFs over a range of 7 magnitudes
($-23 < M_R - 5\ log\ h_{65} < -16$) at both the core and the outskirts 
of the Coma cluster. 
However, in a recent study, Yagi et al (2002) showed that the composite LF
from 10 nearby clusters is well described by a Schechter form while the LFs
for individual clusters are not. Furthermore, they found the
composite LF to have a steeper faint-end slope compared to the general field, 
in contrast to the present study and the results from De Propris et al (2003).
However, this is $<2\sigma$ effect and not statistically significant. 

The characteristic magnitudes of LFs derived here agree with recent
estimates from Beijersbergen et al
($M^\ast_R = -21.91^{+0.26}_{-0.34} + 5\ log~h_{65}$;
$M^\ast_B = -20.03^{+0.36}_{-0.40}  + 5\ log~h_{65}$), after converting 
to the waveband used in this study. 
However, comparison with a large sample of nearby ($z < 0.11$)- 
(De Propris et al 2003) 
and intermediate redshift ($z\sim 0.3$)- (Yagi et al 2002; Valloto et al 1997;
Gaidos 1997)
clusters show that while
$M^\ast_R$ values are consistent ($M^\ast_R = -22.04\pm 0.11 + 5\ log\ h_{65}$
(Gaidos (1997) and $M^\ast_R = -22.24\pm 0.20 + 5\ log\ h_{65}$ 
(Yagi et al (2002)),  our
$M^\ast_B$ estimate for Coma is fainter by almost one magnitude
($M^\ast_B = -20.80 \pm 0.10 + 5\ log\ h_{65}$ (De Propris et al (2003) and 
Valloto et al (1997))).  Differences between the faint-end
slope of the LFs between these studies have only a small effect in this
comparison. This indicates that galaxies in the Coma 
are, on average, redder than other clusters. This is unlikely to be due to a
redshift effect, as De Propris et al sample covers a range $z < 0.11$. Also,
the richness of the Coma cluster cannot be responsible for this result, since
there are also many rich clusters in the Yagi et al sample. 
Moreover, fitting a subset
of their clusters with $\sigma < 800$ Km/sec, De Propris et al still estimate
$M^\ast_B = -20.72 \pm 0.10 + 5\ log\ h_{65}$, similar to that from their
full sample. This is likely due to the very large fraction of early-type
galaxies in the Coma cluster, as also shown in Figure 9.

A dip is detected in the total LFs for the Coma cluster in both R and 
B-bands. This appears at the same magnitude, scaled by the mean $B-R$
color of galaxies in Coma. A similar effect is also found in the LFs
from other studies of the Coma (Biviano et al 1995; Driver et al 1994), 
other nearby (Wilson et al 1997) and intermediate
redshift clusters (Yagi et al 2002; Dahlen et al 2003) but not in deep
field surveys (Ellis et al 1996; Cowie et al 1997). We also find the
amplitude of the dip to increase from R to B-band. The presence of
the dip implies a two-component shape for the cluster LF, consisting of a 
Gausssian and Schechter function at the bright and faint magnitudes 
respectively
(Biviano et al 1995; Yagi et al 2002). Considering the total LF in
$B-R$ color intervals in Figure 9, we find a distinct change in the
shape of the LF for red galaxies ($B-R > 1.35$) at $M_R=-18$ mag., both
at the core (Coma1) and outskirt (Coma2 and Coma3) fields. This result
is in qualitative agreement with the finding by Yagi et al (2002) 
who attribute the change in the
LF shape to contribution from early-type galaxies, defined as those
with $r^{1/4}$-law surface brightness profile. 

The question of particular interest here concerns 
the nature of galaxies contributing to the faint-end of the Coma LF and if
these are different between the Coma1 and Coma3 fields. 
Study of the LF in color intervals (Fig. 9) shows an abundance of
low luminosity red galaxies ($B-R > 1.3$) in Coma 1, compared to star-forming
objects ($B-R < 1.3$), with the difference decreasing towards the outskirts
(Coma2 and Coma3). Moreover, the faint-end
slope of the red component of the LF is significantly steeper 
($\alpha= -1.6$) than that observed for blue galaxies ($\alpha=-1.1$). 
This implies that low luminosity red galaxies
dominate the faint-end slope of the LF, in agreement with the result in
Conselice (2001) who finds a steep faint-end slope for
the LF in A0426 due to the presence of a low-mass red population. 
Therefore, the main contribution to the faint-end of the Coma LF (and
in clusters studied by De Propris et al (2003)), is the large number of
faint red galaxies that are present (presumably produced from the equally
large number of faint blue galaxies in the field). A relatively
higher surface density for this population in the denser region of 
the cluster (Coma1) compared to the less dense region (Coma3) indicates that
dynamical stripping of high mass systems in cluster environments is, at least
partly, responsible for the formation of low-luminosity red galaxies. 
However, 
other mechanisms must be at work in less dense regions of Coma to 
produce this population. Evidence for gas deficiency in dwarf galaxies
in Coma3 comes from the absence of HI emission around N4839 group 
(Bravo-Alfaro 2000). This indicates that low luminosity galaxies in 
the outskirts of the Coma cluster have gone
through a gas-loss process, either through SNe winds (due to their small 
potential well) or by dynamical effects after passing through the cluster. 
Indeed, in a study of radial dependence
of spectroscopic line indices of these galaxies (after correcting for
luminosity dependence), Carter et al (2002) find a significant gredient
in Mg$_2$ , in the sense that galaxies in the core have stronger Mg$_2$
indices and hence, higher metallicities. This is caused by trapping of
the material in galaxies in dense regions due to external
pressure by intracluster medium, leading to higher metallicities. 
Moreover, X-ray morphology
of the N4839 group in Coma3 field indicates that this group is in the
process of infall, showing evidence of interaction with the cluster 
(Neumann et al 2001). Further observational
evidence for dynamical stripping in dense regions of clusters comes
from the discovery of diffuse arcs at the core of the Coma cluster
(Trentham \& Mobasher 1998; West \& Gregg 1999). 

There appears to be a clear separation in luminosity distribution of galaxies
as a function of their effective surface brightness. This monotonic 
surface brightness-luminosity relation
is useful in studying properties of dwarf galaxies with respect to giants
(Ferguson \& Binggeli 1994). Spectroscopically confirmed members of the Coma
cluster have a relatively brighter effective surface brightness 
($\mu_{eff} < 20 $ mag./arcsec$^2$) compared to field galaxies, implying
destruction of low surface brightness galaxies in environments with 
higher local densities. This is predicted by the harassment scenario
of galaxy formation, that due to small potential well of low luminosity/
low surface brightness galaxies, up to 90\% of their stellar content could
be ejected in dense regions of clusters due to interaction with larger
galaxies (Moore et al 1996). However, this same process also 
predicts formation of 
low surface brightness galaxies through stripping at the cores of rich
clusters (Moore et al 1999). Therefore, the present data cannot be used to 
constrain models for the formation of low surface brightness galaxies. High
resolution spectroscopy, measuring features diagnostic of age and metallicity
is found to be more effective in studying formation and evolution of 
low surface brightness galaxies (Carter et al 2002).

\section {Summary}

The spectroscopic survey of galaxies in the Coma cluster, 
performed by Mobasher et al (2002), is compiled with a brighter sample
(Edwards et al 2002),  
providing a wide-area (1 deg$^2$) and deep ($R\sim 19.5$) survey, covering
both the core and outskirts of the cluster. The final survey consists of a
total of 1191 galaxies of which, 760 galaxies are spectroscopically 
confirmed members of the Coma cluster. After correcting the 
spectroscopic sample for incompletness, the LFs are constructed,  
spanning the range $-23 < M_R - 5\ log\ h_{65} < -16$, and covering both
the core and outskirts of the cluster. Results from this study are
summarised below:

\begin{itemize}

\item The R-band LFs are similar at the core and outskirts of the Coma 
cluster, 
with no evidence of a steep faint-end slope, found in previous studies
(mostly based on photometric surveys). However, due to spectroscopic 
limitations, the current sample only extends to $M_R = -16$ mag. while
the observed up-turn in the LF slope is seen at somewhat fainter
magnitudes in photometric surveys. 
The total R-band LF for the Coma cluster, fitted to a Schechter form, is;
$M^\ast_R = -21.79^{+0.08}_{-0.09} + 5\ log\ h_{65}$; 
$\alpha = -1.18^{+0.04}_{-0.02}$. This is found to be close
to that in the general field.

\item As the R-band mostly measures contributions from the old stellar
population, relatively unaffected by star formation, the similarity of
the LFs implies that the average luminosity for a given
galaxy mass is little affected by the processes which dictate their 
morphologies and spectral types.

\item The total B-band LF, fitted to a Schechter form, is:
$M^\ast_B = -19.95^{+0.15}_{-0.10} + 5\ log\ h_{65}$; 
$\alpha = -0.96^{+0.01}_{-0.02}$. This shows a dip at
$M^\ast_B = -18$ mag. A similar feature, although with smaller
amplitude, is observed in the total R-band LF at the same luminosity
(shifted by mean $B-R$ color). This is likely to be due to contribution
from luminous early-type galaxies ($M^\ast_B < -18$ mag.) to the total LF. 
This suggests that the total LF could best be fitted by 2-components, a
gaussian (at bright magnitudes) and a power-law (at fainter magnitudes). 

\item A study of the LFs in $B-R$ color intervals show a steep faint-end
slope for red ($B-R > 1.35$) galaxies, both at the core and the outskirts of
the cluster. This population of low luminosity red galaxies has a higher
surface density than the blue 
($B-R < 1.35$) star-forming population (both at the core and outskirts
of the cluster), and dominates the faint-end of Coma cluster LF. 

\item A monotonic correlation is found between the effective 
surface brightness and luminosity of galaxies. Cluster galaxies are found
to have a higher surface brightness than their field counterparts. 
This implies destruction of low surface brightness galaxies
in dense regions of clusters, indicating effect of 
local environment on formation of galaxies.  

\end{itemize}

{\bf Acknowledgement} We are grateful to an anonymous referee for very useful 
suggestions. We also thank Simon Driver for carefully reading the manuscript
and for useful comments.


\begin{thebibliography}{}

\bibitem[adami]{adam00}
Adami, C., Ulmer, M. P., Durret, F., Nichol, R. C., Mazure, A., Holden B. P., 
Romer, A. K., Savine, C. 2000 A\& A 353, 930

\bibitem[andreon]{and01}
Andreon S. \& Cuillandre, J.-C. 2001 astro-ph 0111528

\bibitem[arnaboldi]{arna02}
Arnaboldi, M. et al. 2002 AJ 123, 760

\bibitem[beigersbergen]{car01}
Beijersbergen, M., Hoekstra, H., van Dokkum, P. G. \& van der Hulst, T. 
2002, MNRAS 329, 385

\bibitem[bernstein]{ber95}
Bernstein, G.M., Nichol, R.C., Tyson, J.A., Ulmer, M.P. \& Wittman, D. 1995
AJ 110, 1507

\bibitem[bernstein]{biv95}
Biviano, A., Durret, F., Gerbal, D. Le Fevre, O., Lobo, C., Mazure, A.
\& Slezak, E. 1995 A\& AS 111, 265

\bibitem[blanton]{bl01}
Blanton M. R. et al 2001 AJ. 121, 2358

\bibitem[brav]{brav00}
Bravo-Alfaro, H., Cayatte, V., van Gorkom J. H., Balkowski, C. 2000 AJ. 119,
580

\bibitem[carter]{car02}
Carter, D. et al. 2002 Ap.J. 567, 772 (paper V)

\bibitem[chiba]{chib94}
Chiba, M., \& Nath, B. B. 1994 ApJ. 436, 618

\bibitem[carter]{col96}
Colless, M. \& Dunn, A. M. 1996 ApJ 458, 435

\bibitem[conselice]{con02}
Conselice, J.C. 2002 ApJ. in press (astro-ph 0205364)

\bibitem[cowie]{cow96}
Cowie, L. L., Songaila, A., Hu, E., \& Cohen, J. G. 1996, AJ, 112, 839

\bibitem[cross]{cros01}
Cross, N. et al 2001 MNRAS 324, 825

\bibitem[dahlen]{dah03}
Dahlen, T., Fransson, C., Ostlin, G., Naslund, M. 2003 MNRAS (submitted)

\bibitem[deproprtis]{dep98}
De propris, R., Eisenhardt, P. R., Stanford, S. A. \& Dickinson, M. 1998
AJ 503, L45

\bibitem[deproprtis]{dep02}
De Propris, R. et al. 2003 MNRAS (submitted)

\bibitem[dekel]{dek86}
Dekel A. \& Silk, J. 1986 ApJ, 303, 39

\bibitem[deriver]{der94}
Driver, S. P., Phillipps, S., Davies, J. I., Morgan, I.,
Disney, M. J. 1994 MNRAS 268, 393

\bibitem[deriver]{der95}
Driver, S. P., Windhorst, R. A., Griffiths, R. E. 1995, ApJ 453, 48

\bibitem[edwards]{edi02}
Edwards, S. A., Colless, M., Bridges, T. J., Carter, D., Mobasher, B., 
Poggianti, B. M. 2002, Ap.J. 567, 178

\bibitem[efstathiuo]{efs92}
Efstathiou, G. 1992 MNRAS 256, 43P

\bibitem[ellis]{ell96}
Ellis, R. S., Colless, M., Broadhurst, T. J., Heyl, J. \& Glazebrook K.
MNRAS, 280, 235

\bibitem[ferguson]{fer94}
Ferguson, H. C. \& Binggeli, B. 1994 A \& ARv 6, 67

\bibitem[geidos]{geid97}
Gaidos, E. J., 1997 AJ 113, 117

\bibitem[godwin]{godw83}
Godwin, J. G., Metcalfe, N., Peach, J. V. 1983, MNRAS, 202, 113 

\bibitem[greg]{gre98}
Gregg M. D. \& West, M. J. 1998 Nature 396, 549

\bibitem[komiyama]{kom02}
Komiyama Y., et al. 2002, Ap.JS., 138, 265 (paper I)

\bibitem[lobo]{lob97}
Lobo, C., Biviano, A., Durret, F., Gerbal, D., Le Fevre, O., Mazure, A.,
\& Slezak, E. 1997 A\&A 317, 385

\bibitem[lopez]{lop97}
Lopez-Cruz, O., Yee, H.K.C., Brown, J.P. Jones, C., Forman W. 1997, ApJ, 
475, L97

\bibitem[lin]{lin96}
Lin, H., Kirshner, R. P., Shectman, S. A., Landy, S. D., Oemler, A., 
Tucker D. L., Schechter, P. L. 1996, Ap. J., 464, 60

\bibitem[loveday]{lov92}
Loveday, J., Peterson, B. A., Efstathiou, G. \& Maddox, S. J. 1992, Ap.J. 
390, 338

\bibitem[mateo]{mat98}
Mateo, M. L. 1998 ARA\&A 36, 435

\bibitem[mobasher]{mob01}
Mobasher, B., et al. 2001 ApJS 137, 279 (paper II)

\bibitem[mobasher]{mob98}
Mobasher, B. \& Trentham, N. 1998 MNRAS 293, 315

\bibitem[moore]{moore96}
Moore, B., Katz, N. Lake, G., Dressler, A. \& Oemler, A. 1996 Nature 379, 613

\bibitem[moore2]{moore99}
Moore, B., Lake, G., Stadel, J. \& Quinn, T 1999 {\it ASP Conf. ser. 170}:
The low surface brightness Universe, 229

\bibitem[neu]{neu01}
Neumann, D. M. et al. 2001 A \& A 365, L74

\bibitem[norberg]{nor02}
Norberg, P. et al. 2002 MNRAS in press

\bibitem[phillipps]{phil98}
Phillipps, S., Driver, S.P., Couch, W.J. \& Smith R.M. 1998 ApJ 498, L119

\bibitem[poggianti]{pogi01}
Poggianti, B. M. et al 2001, ApJ 562, 689 (paper III)

\bibitem[rines]{rines01}
Rines, K., Geller, M.J., Kurtz, M. J., Diaferio, A., Jarrett, T. H. \& 
Huchra, J. P. 2001 ApJL 561, 41

\bibitem[schechter]{sch76} 
Schechter, P. L. Ap.J. 1976 203, 297

\bibitem[secker]{sec96}  
Secker, J. \& Harris, W. E. 1996, APJ 469, 628

\bibitem[sek]{sek98} 
Sekiguchi, M., Nakaya, , H., Kataza, H. \& Miyazaki, S., 1998, Exp. Astron.,
8,15

\bibitem[somerville]{som01}
Somerville, R. S., 2001 Ap.J. astro-ph 0107507

\bibitem[sprayberry]{sper97}
Sprayberry, D., Impey, C. D., Irwin, M. J. \& Bothun, G. D. 1997, 482, 104

\bibitem[thompson]{tho93}  
Thompson L. A. \& Gregory, S. A. 1993, AJ 106, 2197

\bibitem[tou]{tou96}
Thoul A. A. \& Weinberg, D. H. 1995, ApJ., 442, 480

\bibitem[trentham]{ter98}
Trentham, N. 1998 MNRAS 293, 71  

\bibitem[ter]{ter98}
Trentham, N. \& Mobasher, B. 1998 MNRAS 293, 53

\bibitem[trentham]{ter01}
Trentham N., Tully, R. B. \& Verheijen M. A. W. 2001 MNRAS 325, 385

\bibitem[trentham]{ter02}
Trentham, N. \& Hodgkins 2002 MNRAS (in press) astro-ph 0202437

\bibitem[tully]{tul01}
Tully, R. B., Somerville, R. S., Trentham, N. \& Verheijen, M. A. W. 
2001 ApJ astro-ph 0107538

\bibitem[val1]{val97}
Valotto, C. A., Nicotra, M. A., Muriel, H., \& Lambas, D. G. 1997 Ap.J. 
479, 90

\bibitem[val2]{val01}
Valotto, C. A., Moore, B., Lambas, D. G. 2001 Ap. J., 546, 157

\bibitem[west]{west00}
West, M. \& Blakeslee, J. P. 2000, ApJ 543, L27

\bibitem[wilson]{wils97}
Wilson, G., Smail, I., Ellis, R. S. \& Couch, W. J. 1997 MNRAS 284, 915

\bibitem[yagi]{yag02} 
Yagi, M., Kashikawa, N., Sekiguchi, M., Doi, M, Yasuda, N, Shimasaku, K
\& Okamura, S. 2002 Ap.J. 123, 87

\bibitem[yasuda]{yas97}
Yasuda, N., Fukugita, M. \& Okamura, S. 1997 ApJS 108, 417

\end{thebibliography}
\end{document}